\documentstyle[12pt]{article}
\newcommand{\doublespace}{
   \renewcommand{\baselinestretch}{1.75}
   \large\normalsize}
\newcommand{\be}{\begin{equation}}
\newcommand{\ee}{\end{equation}}
\begin{document}
\baselineskip0.85cm
\title{Kondo Insulators Modeled by the
One Dimensional Anderson Lattice: A Numerical Renormalization Group Study}
\author {M.Guerrero and Clare C. Yu \\ 
Department of Physics\\ University of California, Irvine\\ Irvine, CA 92717}
\date{ \ \ \ }

\maketitle
\setcounter{page}{0}
\thispagestyle{empty}
\begin{abstract}
\doublespace

In order to better understand Kondo insulators,
we have studied both the symmetric and asymmetric Anderson 
lattices at half-filling in one dimension using the density 
matrix formulation of the numerical 
renormalization group. The asymmetric case is treated in the
mixed valence regime.
We have calculated the charge gap, spin gap and quasiparticle gap as a 
function of the repulsive interaction U using open boundary conditions
for lattices as large as 24 sites. 
We find that the charge gap is larger than the spin gap for all U for both
the symmetric and asymmetric cases. RKKY interactions are evident in the 
f-spin-f-spin correlation functions at large U in the symmetric case, but are 
suppressed in the asymmetric case as the f-level approaches the Fermi energy.
This suppression can also be seen in the staggered susceptibility 
$\chi(q=2k_{f})$, and it is consistent with neutron scattering measurements
of $\chi(q)$ in CeNiSn. 
\end{abstract}


\newpage

\doublespace
\section{Introduction}

Kondo insulators are a class of rare earth compounds that become semiconducting at
low temperatures. However, they
are not just ordinary semiconductors because
their gaps are due to many body interactions between extended
conduction and localized electrons in the f-shell.
At high temperatures ($T > 100$K), these materials
behave as metals with local magnetic moments, but at low temperatures
they exhibit a small gap in their excitation spectrum and behave
as narrow gap semiconductors (typically $\Delta\sim 100$K).
This class of materials includes 
Ce$_{3}$Bi$_{4}$Pt$_{3}$,
CeNiSn, and SmB$_{6}$ \cite{Fisk}. It has also been argued
that the transition metal compound, FeSi, is also a Kondo
insulator \cite{Fisk}.

Theoretically, the Kondo insulators have been modeled by
both the Kondo lattice and the periodic Anderson model
at half filling \cite{hiro,zqw,Yu,Fye,Nishino,Blankenbecler}.
In the Anderson lattice each site has a localized orbital
that hybridizes with an extended band of conduction electrons.
Double occupation
of the localized orbital is penalized by a strong Coulomb repulsion U.
The half-filled system,
in which there are two electrons per site, is an insulator
at zero temperature. 

In the Anderson model, two distinct regimes can be considered.
In the {\em Kondo regime}, the energy of the localized band lies well
below the Fermi surface. In this case,
the occupation of the localized level is very close to one, and
charge fluctuations in and out of the localized orbital
are negligible. 
On the other hand, in the {\em mixed valence} regime,
the energy of the localized
orbital lies very close to the Fermi level, and electrons
can hop in and out of the localized orbital.
As a result, charge fluctuations are important and
the average occupancy of the localized orbital is less than one.

Most of the theory that has been done on Kondo insulators has
treated them in the Kondo regime or assumed particle-hole symmetry
or both \cite{hiro,zqw,Yu,Fye,Nishino,Blankenbecler}. 
(The symmetric Anderson lattice has particle-hole symmetry while
the asymmetric case does not.)
However, real materials do not necessarily have particle-hole symmetry,
and it is generally believed that
they are more likely to be in the mixed valence
regime. Varma has argued that charge fluctuations
are crucial to the understanding of the physics of
these systems $\cite{Varma}$. He points out that such
fluctuations should reduce the magnetic correlations between
sites. As we shall see in section 4.2, 
this is confirmed by our calculations.
Indeed neutron scattering experiments have indicated
the absence of magnetic correlations in CeNiSn $\cite{Mason}$.
In addition, measurements of the occupation of the f-orbital in
$Ce_{3}Bi_{4}Pt_{3}$ give $n_{f}=0.865$ at $T=0$ \cite{Kwei}, 
indicating a mixed valence state.
Thus it is important to determine the properties of the
asymmetric Anderson lattice in the mixed valence regime. 

Previous studies of the mixed valence case have been hampered by
various limitations. Analytic calculations, which include the Gutzwiller
approach \cite{Rice,Varma1} and the slave boson mean field approximation
\cite{Riseborough,millislee}, do not distinguish 
between the different kinds of gaps (spin,
charge and quasiparticle gap). Numerical approaches have been
limited to 4-site chains \cite{Jullien, Callaway} and did not
consider the charge gap \cite{Jullien}. 

In this paper we present the first systematic study of 
all three gaps of the half filled Anderson lattice
in both the Kondo and the mixed 
valence regime. To the best of our knowledge, this is the first time 
that the staggered susceptibility $\chi(q=2k_{f})$, 
the f-spin-f-spin correlation functions, 
and the occupation $n_f$ of the
localized orbital have been calculated for
the mixed valence case.

The density matrix renormalization group technique allows us
to consider chains of up to 24 sites. This gives us an advantage over
other numerical approaches \cite{Fye,Jullien,Callaway} which can only
deal with short chains. 
Not only are we able to get f-spin-f-spin correlation 
functions, but by considering long chains, we can also
study the behavior of the f-spin-f-spin correlations in both regimes.
We establish that the amplitude of the correlations decays exponentially 
in space,
and we can calculate the correlation length as a function of 
the parameters. Previous Monte Carlo studies of the symmetric case
were unable to distinguish between power law or exponential \cite{Fye}
because of the length of the lattice.
In the mixed valence case, where the accuracy is higher than for
the symmetric case at large U, we can
get the extrapolated value of the spin gap and the average
occupation of the
localized orbital for infinite chains. We find that both have 
power law behavior for large U. This analysis was not possible in previous 
studies of short chains \cite{Jullien,Callaway}. 

We also consider the $U=\infty$ in which the doubly occupied states
of the localized orbital are suppressed. 
Several approximation schemes, such as
the Gutzwiller technique and the slave boson mean field approach,
have been applied in this limit.
However, it is not clear how well these approximations describe
the properties of the
Anderson Hamiltonian. To find out, we compare some of 
our results with those of the
slave boson mean field approach. We find good agreement when the system is
in the mixed valence regime but as the system approaches the
Kondo regime, the slave boson picture breakdowns.

The paper is organized as follows. 
We consider the one dimensional
Anderson lattice at half filling using the
density matrix formulation of the quantum
renormalization group technique \cite{White,White1} which is
briefly described in section 2. After discussing
the Anderson and Kondo lattice Hamiltonians in section 3,
we give our results in section 4. In both the symmetric and asymmetric case, 
we find in section 4.1 that the charge gap is larger than the spin gap.
In section 4.2, we study the RKKY interactions
by calculating the f-spin-f-spin correlation function
and the staggered susceptibility
$\chi(q=2k_{f})$ as a function of $U$. 
Our results show that
in the Kondo regime the RKKY interactions
become important as $U$ increases,
while in the mixed valence case such magnetic
correlations are suppressed because the f-orbitals are
not always occupied. In section 4.3, we find that
the occupation $n_f$ of the localized orbital
changes from 1 at $U=0$ to a value close to 0.7 at 
$U=\infty$ in the asymmetric case when the localized band
is at the Fermi energy. (In the symmetric case
$n_f=1$.) In section 4.4, we study the $U=\infty$ case. 
We present our conclusions in section 5.

\section{Quantum Renormalization Group Technique}

The density matrix formulation of the numerical renormalization group has
proven extremely useful for the study of the ground state and low energy excitations of
one-dimensional systems \cite{White2,Sorensen}. 
It has distinct advantages over other computational
techniques. For example, it allows us to  study chains longer than what is achievable by
exact diagonalization, and it avoids the problem of Monte Carlo calculations where
very low temperatures are difficult to attain.

In real space renormalization group schemes, the system is divided into
small blocks. The traditional renormalization group
approach consists of diagonalizing the Hamiltonian of a small block,
keeping the lowest energy eigenstates, and then forming a bigger block
by combining a few of the small blocks.
The drawback of this approach is that it
completely neglects quantum fluctuations at the boundary of the blocks. 
For this reason, the standard approach performs poorly and 
there is not a significant improvement 
by increasing the number of states kept $\cite{White}$. 

The density matrix formulation
of the renormalization group provides a systematic method of choosing the states to keep for
the bigger block and at the same time it takes into account the quantum fluctuations 
at the boundaries.
The procedure for a one dimensional lattice goes as follows. 
We divide the chain into four blocks. Typically, blocks 1 and 4 
have many sites and blocks 2 and 3 (in the center of the chain) are single sites. 
We find the ground state of the
Hamiltonian for the lattice:
\begin{equation}
|\psi> = \sum_{i1,i2,i3,i4} \psi_{i1,i2,i3,i4} | i1, \,\,i2, \,\, i3, \,\, i4>
\end{equation}
where $i1$, $i2$, $i3$, and $i4$ label states in blocks 1, 2, 3 and 4.
Then we consider blocks 1 and 2 to be the ``system'' 
and the rest of the chain to be the ``environment''. 
The objective is to construct a new block 1 out of blocks 1 and 2 by keeping only a fraction
of the total number of states. In order to choose which states to keep, 
we form the density matrix for 
blocks 1 and 2 by tracing out the degrees of freedom of blocks 3 and 4 $\cite{Feynman}$:
\begin{equation}
\rho(i1,i2,ii1,ii2)=\sum_{i3,i4} \psi_{i1,i2,i3,i4} \cdot \psi^{*}_{ii1,ii2,i3,i4}  
\end{equation}
We then diagonalize this density matrix. The eigenvalues of the density matrix 
give the weight of the associated eigenstates $\cite{Feynman}$. 
Since it is not possible to 
keep all the states, we truncate the basis by keeping 
a predetermined number of states with the largest weights. 
We use these eigenstates to construct the basis of the new block 1 which
is made out of the old blocks 1 and 2. In this way, we increase the
size of block 1 by one site.

It is possible to target more than just the ground state; the lowest excited states 
can be calculated as well by simply diagonalizing the Hamiltonian for these
states. In each case the diagonalization is performed 
in a subspace with fixed quantum numbers, e.g., with a fixed number of electrons 
and a fixed z-component of the
total spin. 
This method performs at its best for open boundary conditions in which 
there is no hopping past the ends of the chain. 
We typically keep 
100 states, although for cases where higher accuracy is needed, we keep up to 200 states. 
For small values of the Coulomb repulsion, the results are very accurate 
with truncation errors of the
order of $10^{-8}$, while for large values of the Coulomb repulsion  the accuracy 
is reduced and the 
truncation errors increase to $10^{-4}$.
For a more detailed description of the method, see reference $\cite{White}$.

As a check for our method, we compare our energies and 
correlation functions with exact diagonalization results. For short chains
we can keep all the states, and we obtain full agreement with exact diagonalization.
As another check, we compare the ground state energy for the
symmetric Anderson lattice with Gutzwiller $\cite{Gulacsi}$ and Monte Carlo results 
$\cite{Blankenbecler}$ as shown in Fig. 1. We plot $(e_{o}+U/2)/|e_{o}|$, 
where $e_{o}$ is the
ground state energy per site. The Monte Carlo simulations 
calculated the ground state energy 
for 16 site chains
with periodic boundary conditions $\cite{Blankenbecler}$. We use open boundary 
conditions and chains of 24 sites. The Gutzwiller values are for a
correlated spin density wave $\cite{Gulacsi}$. We obtain 
good agreement with both calculations. For large U, our energies 
lie below the Monte Carlo and the Gutzwiller 
values, giving a better upper bound for the ground state energy. In this regime
the binding energy is proportional to $1/U$ in agreement with perturbation 
theory. 

\section{The Periodic Anderson Hamiltonian}

We consider the standard periodic Anderson Hamiltonian in one dimension:
\begin{equation}
H=-t\sum_{i \sigma} (c^{\dagger}_{i \sigma}   c_{i+1 \sigma} +
                     c^{\dagger}_{i+1 \sigma} c_{i\sigma} )
  + \epsilon_{f} \sum_{i \sigma} n^{f}_{i \sigma} 
  + U \sum_{i} n^{f}_{i \uparrow}n^{f}_{i \downarrow}
  + V \sum_{i \sigma} (c^{\dagger}_{i \sigma} f_{i \sigma} + 
                      f^{\dagger}_{i \sigma}  c_{i \sigma} )     \label{eq:Hamiltonian}
\end{equation}
where $c^{\dagger}_{i \sigma}$ and $c_{i\sigma}$ create and annihilate conduction
electrons with spin $\sigma$ at lattice site $i$, $f^{\dagger}_{i \sigma}$ and
$f_{i \sigma}$ create and annihilate local f-electrons,
t is the hopping matrix element for conduction electrons between neighboring sites, 
$\epsilon_{f}$ is the energy of the localized f-orbital,  
U is the on-site Coulomb
repulsion of the f-electrons, and V is the on-site hybridization matrix element
between electrons in the f-orbitals and the conduction  band.
For simplicity we neglect orbital degeneracy. U, V, t, and $\epsilon_{f}$ are 
taken to be real numbers.

Let us examine the Anderson Hamiltonian in various limits.
For $U=0$ this Hamiltonian can be exactly diagonalized in k-space. We obtain
two hybridized bands with energies $\lambda_{k}^{\pm}$:
\begin{equation}
\lambda_{k}^{\pm}=\frac{1}{2}\left[(\epsilon_{f}-2t\cos(ka)) \pm
\sqrt{(\epsilon_{f}+2t\cos(ka))^{2}+4V^{2}}\right]
\end{equation}
where $a$ is the lattice constant.
When there are two electrons per unit cell, the lower band is full while the upper one
is empty. Thus the system is insulating when $N_{e}=2L$. Here $N_{e}$ is the number
of electrons and $L$ is the number of sites in the lattice. 


When the mixing term is small 
($\pi V^{2}/2t(\epsilon_{f}+U) \ll 1$ and
$\pi V^{2}/2t\epsilon_{f} \ll 1 \,$), 
the Anderson Hamiltonian can be mapped into 
the Kondo model $\cite{Schrieffer}$
\begin{equation}
H_{K}=-t\sum_{i \sigma} (c^{\dagger}_{i \sigma}   c_{i+1 \sigma} +
                     c^{\dagger}_{i+1 \sigma} c_{i\sigma} ) +
J_{eff}\sum_{i} \vec{S}^{f}_{i} \cdot \vec{s}^{c}_{i} 
\end{equation}
where $\vec{s}^{c}_{i}$ is the spin density of the conduction electrons at site $i$ and
$J_{eff}$ is given by the Schrieffer-Wolff transformation $\cite{Schrieffer}$:
\begin{equation}
J_{eff}=-\frac{2|V|^{2}U}{\epsilon_{f}(\epsilon_{f}+U)} \label{eq:Jeff}
\end{equation}
Note that for the symmetric case where $\epsilon_{f}=-U/2$, $J_{eff}=8V^{2}/U$, and for
$U=\infty$, $J_{eff}=-2V^{2}/\epsilon_{f}$. 

Although there is no direct interaction between the f-electrons 
in the Hamiltonian (~\ref{eq:Hamiltonian}), there is an indirect 
exchange coupling between the f-sites via the conduction band.  
This is the RKKY coupling, originally derived for the magnetic interaction 
between nuclei in metals $\cite{Ruderman}$. An electron in the f-orbital 
tends to polarize the spin of the conduction electron cloud around it. 
The spin of the conduction electron polarization cloud oscillates with distance
and it polarizes the spin of the f-electrons at neighboring sites. 
In the case of the Anderson Hamiltonian, the RKKY coupling constant 
can be obtained by considering
two Anderson impurities in a gas of conduction electrons. 
The hybridization term is taken as 
a perturbation, and the fourth-order contribution to the perturbation 
expansion gives the coupling between sites.
The effective Hamiltonian for the two impurities is:
\begin{equation}
H_{RKKY}=J_{RKKY}\vec{S_{1}^{f}} \cdot \vec{S_{2}^{f}} \label{RKKYH}
\end{equation}
where $\vec{S_{1}^{f}}$ and $\vec{S_{2}^{f}}$ are the spin 
operators for the f-electrons on the two impurities.
The explicit expression for $J_{RKKY}$ is fairly complicated in general \cite{Fye}, 
but it can be shown that it favors antiferromagnetic alignment of the 
f-spins with wavevector $q=2k_{f}$, where $k_{f}$ is the Fermi
wavevector of the noninteracting conduction electrons. 

For the one dimensional symmetric case ($\epsilon_{f}=-U/2$) at half filling, 
$2k_{f}a=\pi$, and the asymptotic expression
for large distances is
$\cite{Fye}$
\begin{equation}
J_{RKKY}=\left( \frac{8V^{2}}{U}\right)^{2}\frac{1}{8\pi t} \frac{(-1)^l}{l} \label{RKKY1}
\end{equation}
where $l$ is the distance between $\vec{S_{1}}$ and $\vec{S_{2}}$ 
in units of the lattice constant.
Note that $J_{RKKY}=(-1)^{l}J_{eff}^{2}/8\pi tl$, that is,
the Kondo coupling constant is the parameter that governs the f-f correlations in the Kondo 
regime. The asymptotic expression (~\ref{RKKY1}) is the same as the one 
obtained by 
starting with the Kondo Hamiltonian and doing second-order perturbation theory in $J_{eff}$.

We expect RKKY interactions to be important in the Kondo regime where the occupancy
of the f-orbital is very close to 1. In the mixed valence regime, the occupancy of
the f-level is reduced, and therefore RKKY correlations should be suppressed
\cite{Varma}.

\subsection{Symmetric case}

The symmetric case corresponds to a value of $\epsilon_{f} =-U/2$. 
For large U, the f-level lies well 
below the Fermi surface and the system is in the Kondo regime. 
Particle-hole symmetry ensures that the occupancy of the f-level, 
$n_{f}, $ is always one. 
In the strong coupling regime ($8V^{2}/U \ll t$), charge fluctuations in the f-orbital are highly 
suppressed because the Coulomb repulsion makes the states with one electron in the 
f-level much more energetically favorable 
than the state with two electrons or the state with the orbital empty.

In addition
 to the usual SU(2) spin symmetry, this
particular choice of $\epsilon_{f}$ introduces an SU(2) 
charge pseudospin symmetry into the system
 $\cite{Nishino}$.
The pseudospin operator ($\vec{I}$) can be obtained 
from the spin operator $\vec{S}$ by performing a particle-hole
transformation on one of the spin species \cite{Nishino}. Its components are:
\begin{eqnarray}
\nonumber  I_{z} & = & \frac{1}{2} \sum_{i} (c^{\dagger}_{i \uparrow}c_{i \uparrow}+
	    c^{\dagger}_{i \downarrow}c_{i \downarrow}+
	    f^{\dagger}_{i \uparrow}f_{i \uparrow}+
            f^{\dagger}_{i \downarrow}f_{i \downarrow} -2) \\
            I_{+} & = & \sum_{i} (-1)^{i} (c^{\dagger}_{i \uparrow}
                                          c^{\dagger}_{i \downarrow}-
	     f^{\dagger}_{i \uparrow}f^{\dagger}_{i \downarrow}) \\
\nonumber  I_{-} & = & \sum_{i} (-1)^{i} (c_{i \downarrow}c_{i \uparrow}-
	f_{i \downarrow}f_{i \uparrow})  
\end{eqnarray}
The z component of the pseudospin is equal to $N_{el}/2 - L$. Note that
half filling corresponds to $I_{z}=0$. An $I_{z}=1$ state can be achieved by
adding two electrons.

All the energy 
eigenstates of the symmetric Anderson model have a 
definite value of S and I. At half filling
the ground state is a singlet both in spin and pseudospin space 
$(S=0, I=0)$ \cite{Ueda}. The lowest lying excited state is
a spin triplet. The energy difference between these two states is
the spin gap $\Delta_{s}$:
\begin{equation}
\Delta_{s} =  E(S=1, I=0)-E_{o}(S=0, I=0)
\end{equation}
where $E_{o}$ is the energy of the ground state.

To find the charge gap, we note that optical experiments measure the
charge gap by measuring the conductivity which is determined by the
current-current correlation function. The current is related
to the charge density through the continuity equation. Thus the lowest
lying charge excitation is the lowest excited state $|n>$ with $S=0$
such that  $<0|\sum_{q} \rho_{q}|n>\neq 0$, where $\rho_{q}$ is the q-component of
the Fourier transformed charge density operator and $|0>$ is the ground state
\cite{millis}. Notice that $\rho_{q}$ is related to $\vec{I}^{z}_{q}$,
where $\vec{I}_{q}$ is a Fourier transformed vector in pseudospin 
space given by
\begin{eqnarray}
\nonumber  I^{z}_{q} & = & \frac{1}{2} \sum_{i} e^{-i\vec{q}\cdot \vec{r}_{i}}
	    (c^{\dagger}_{i \uparrow}c_{i \uparrow}+
            c^{\dagger}_{i \downarrow}c_{i \downarrow}+
            f^{\dagger}_{i \uparrow}f_{i \uparrow}+
            f^{\dagger}_{i \downarrow}f_{i \downarrow} -2) \\
            I^{+}_{q} & = & \sum_{i} e^{-i\vec{q}\cdot \vec{r}_{i}} 
	 		(-1)^{i} (c^{\dagger}_{i \uparrow}
                                          c^{\dagger}_{i \downarrow}-
             f^{\dagger}_{i \uparrow}f^{\dagger}_{i \downarrow}) \\
\nonumber  I^{-}_{q} & = & \sum_{i} e^{-i\vec{q}\cdot \vec{r}_{i}}
	(-1)^{i} (c_{i \downarrow}c_{i \uparrow}-
        f_{i \downarrow}f_{i \uparrow})
\end{eqnarray}
Using the Wigner-Eckart theorem, one can show that
the (I=1, S=0) states are the only states
$|n>$ for which the
charge density $\rho_{q}$
has finite matrix elements $<n|\rho_{q}|0>$ with the
ground state $|0>$. Thus
the charge gap $\Delta_{c}$ is the energy difference between the
ground state and the lowest pseudospin triplet state $\cite{Nishino}$:
\be
\Delta_{c}= E(S=0, I=1)-E_{o}(S=0, I=0)
\ee

The quasiparticle gap gives the energy for making a noninteracting particle and hole.
It is defined as follows:
\be
\Delta_{qp} = E_{o}(N_{e}=2L+1)+E_{o}(N_{e}=2L-1)-2E_{o}(N_{e}=2L) 
\label{eq:qpgap}
\ee
Due to the particle-hole symmetry of the symmetric case, one can write
\be
\Delta_{qp}=2[E_{o}(N_{e}=2L+1)-E_{o}(N_{e}=2L)]
\ee
The quasiparticle gap can be thought of as the difference of chemical
potentials
\be
\Delta_{qp} = \mu_{N+1} - \mu_{N}
\ee
where $\mu_{N}=E_{o}(N_{e})-E_{o}(N_{e}-1)$ and $N_{e}=2L$ for half filling.

For $U=0$, all the gaps coincide and are given by the hybridization gap:
\begin{equation}
\Delta_{s}=\Delta_{c}=\Delta_{qp}=\lambda_{0}^{+}-\lambda_{\pi}^{-}=2\sqrt{t^{2}+ V^{2}} - 2t
\end{equation}
As $U \rightarrow \infty$, the f-electrons decouple from the conduction electrons, 
and all the gaps go to zero.

We can also define the gap $\Delta_{ns}$ between the ground state and the lowest
excited neutral singlet state which has quantum numbers (S=0, I=0) \cite{Yu}.
For the half filled Kondo lattice, the neutral singlet has been found to be
an elementary excitation consisting of a ``particle'' and a ``hole''
which are (S=1/2, I=1/2) excitations. 
In a single site basis, a ``hole'' is a site with one f-electron and no conduction electrons
with quantum numbers (S=1/2, I=1/2, I$_{z}=-1/2$), while a ``particle'' is a site with one
f-electron and two conduction electrons with (S=1/2, I=1/2, I$_{z}=+1/2$).
The Kondo and Anderson lattices agree for 
small $J_{eff}/t$. In this regime the neutral singlet gap is smaller 
than the charge and quasiparticle
gaps but larger than the spin gap \cite{Yu}. However, for small $U$, 
the lowest lying (S=0, I=0) excited state is not an elementary excitation,
but presumably consists of two spin excitations. Indeed, for $U=0$, one can show
that $\Delta_{ns}=2\Delta_{s}$. In the small $U$ limit, there are
bands of spin and charge excitations which lie between the lowest excited neutral singlet
state and the ground state. Because of this, it is hard for
the numerical renormalization group technique to accurately calculate the
energy of the neutral singlet state for small $U$. For this reason, 
we have not calculated the neutral singlet gap.

\subsection{Asymmetric case}

The symmetric case is typically studied in the strong coupling regime 
($8V^{2}/U \ll t$) where the f-level lies far below
the Fermi energy. However, in real materials the f-level can be near
the Fermi surface even though $U$ is large. In addition real materials
ordinarily do not have particle-hole symmetry. For this 
reason we consider the asymmetric 
case in which $\epsilon_{f}$ can have any value irrespective of U. In particular, we place the 
f-orbital right at the Fermi level ($\epsilon_{f}=0$) so that 
states with no electrons 
and states with one electron in the f-level are equal in energy, while states
with two electrons are highly suppressed due to the strong Coulomb repulsion. 
In this mixed valence case, charge fluctuations are allowed and 
the occupation of the f-level, $n_{f}$, is always less than one for $U>0$. 

Note that there is no particle-hole symmetry 
and the pseudospin operator is no longer conserved. Only the
total number of electrons ($I_{z}$) remains a good quantum number. 
For all the values of the parameters that we explored, we found 
that the ground state is still a spin singlet at half filling. 
There is still a gap to the lowest lying excited state which is a spin triplet.
Thus the spin gap can still be defined as:
\begin{eqnarray}
\Delta_{s}=E(S=1)-E_{o}(S=0)
\end{eqnarray}
The quasiparticle gap is defined in equation (~\ref{eq:qpgap}). Note
that in this case $E(2L+1)\neq E(2L-1)$ since there is no particle hole symmetry.
 
Unlike the symmetric case we can no longer use charge pseudospin symmetry
to find the charge gap. We must use the fact that the lowest
lying charge excitation is the lowest excited state $|n>$ with $S=0$
such that  $<n|\sum_{q} \rho_{q}|0>\neq 0$, where $\rho_{q}$ is the q-component of
the Fourier transformed charge density operator and $|0>$ is the ground state
\cite{millis}. Using the fact that $\sum_{q} \rho_{q}=c^{\dagger}_{i=0}c_{i=0}$,
we find that $|n>$ is the lowest $S=0$ excited state. Let its energy
be denoted by $E_{1}(S=0)$. Then the charge gap is 
given by 
\begin{eqnarray}
\Delta_{c}=E_{1}(S=0)-E_{o}
\end{eqnarray}

\section{Results}

In using the density matrix renormalization group technique, we consider lattices
up to 24 sites long, and we use open boundary conditions. We set $t$=1 so that
energies are measured in units of $t$. 
For the symmetric case, by definition $\epsilon_{f}=-U/2$. In order
to study the asymmetric case in the mixed valence regime,
we set $\epsilon_{f}=0$, that is, 
$\epsilon_{f}$ is at the Fermi energy. 
We fix V=1 and vary U from 0 to 30. 
Although this value of V is somewhat larger than in 
real materials, it is convenient for numerical reasons and for comparison with
previous calculations $\cite{Nishino}$.

\subsection{Gaps}

In Figs. 2a and 2b, we plot the gaps versus U for
different chain sizes for the symmetric and asymmetric cases, 
respectively. 

{\it Relative Gap Sizes.}
In both the symmetric and asymmetric cases, the
charge and quasiparticle gaps are larger than the spin gap
for $U>0$. This is consistent with experiments which find that the
charge gap is larger than the spin gap in Ce$_{3}$Bi$_{4}$Pt$_{3}$
\cite{bucher}. As we shall see, in the mixed valence case the ratio
$\Delta_{c}/\Delta_{s}
\sim 2$ for large $U$. This agrees with the experimental
value of 1.8 for Ce$_{3}$Bi$_{4}$Pt$_{3}$ \cite{bucher}.
In the Kondo regime,
one expects a much larger ratio when U is large; in fact,
this ratio diverges in the symmetric case as $U\rightarrow\infty$ 
\cite{Nishino}.
 
Note that the gaps have a smaller value at $U=\infty$
than at $U=0$. In the asymmetric case this
reduction is not as large as in the symmetric case.
This decrease is due to many body interaction
effects and may explain why band structure calculations \cite{Fu,Mattheiss}
find a larger gap than the optical gap measured experimentally
\cite{schlesinger}.

{\it Nature of the Excitations.}
For the symmetric case (Fig. 2a), notice that the spin gap decreases
monotonically with increasing U.
In contrast to the monotonic behavior of the spin gap, 
Fig. 2a shows that
the charge and quasiparticle gaps initially increase with
U, go through a maximum and then decrease. This behavior was also observed
by Steiner {\it et al.} $\cite{Steiner}$. 
To understand this, note that the quasiparticle gap is obtained by adding a particle to
the system. 
For very small $U$, this particle goes predominantly into an f-orbital; so as U 
increases, the gap also increases due to the Coulomb repulsion. 
However, for large
$U$, the extra particle goes mostly into the conduction band, and the gap
starts decreasing. 
One can see how much of an additional particle
goes into an f-orbital by plotting $<N_{f}>-L$ vs. U for the state with
$N_{el}=2L+1$ (see Fig. 3). Since the total number of f-electrons,
$N_{f}$, equals L at half-filling with particle-hole symmetry,
$<N_{f}>-L$ tells us how much of
the extra particle is in the f-orbital.
One can see in Fig. 3 that the extra particle goes mainly into an f-level for
small U but, as U increases, it goes more and more into the conduction band.

Now we discuss the charge gap shown in Fig. 2a. In the symmetric case,
the charge gap is obtained by adding two particles to the 
half-filled system to get an $(I=1, I_{z}=1)$ state. 
It has a maximum as a function of $U$ for 
the same reason that the quasiparticle gap does. The fact that
$\Delta_{c} > \Delta_{qp}$ for any finite value of U means that the
two extra particles repel each other. So as $L \rightarrow \infty$, they
will be infinitely apart
and $\Delta_{c}$ will equal $\Delta_{qp}$.

In Fig. 2b we show the gaps versus U in the asymmetric case. 
We see that for $U \stackrel{<}{\sim} 2,\, \Delta_{c}$
is greater than $\Delta_{qp}$,
but as U increases, they cross.
This crossover can be interpreted as follows. The quasiparticle gap
consists of two noninteracting $S=1/2$ excitations.
The charge gap, on the other hand, is given by the
lowest $S=0$ excited state. For small U this state consists of
two $S=1/2$ excitations, while for large U there is a crossover to a state
made out of two $S=1$ excitations. This picture is supported by the fact that
for small U, $\Delta_{c} \sim \Delta_{qp}$, 
but as U increases, $\Delta_{c}\sim 2\Delta_{s}$.

{\it Finite Size Effects as $U \rightarrow \infty$.}
As $U \rightarrow \infty$ in the symmetric case, the conduction electrons and
the f-electrons decouple, and the gaps approach their values
for a free electron band
on a finite size lattice with open boundary conditions.
We confirmed this by calculating
the gaps on finite size lattices for free electrons.
The charge and quasiparticle gaps have large finite size effects,
especially for large $U$.
Nishino and Ueda \cite{Nishino}
find a much smaller charge gap in the large U limit
because they use periodic boundary conditions, and the charge gap goes
to zero for free electrons on a ring.
On the other hand, our values for the spin gap agree with the
results of Nishino and Ueda $\cite{Nishino}$ because 
the spin gap has small finite size effects.
To understand this, note that the
the spin excitation has small dispersion
because it is a spin flip that has mainly f-character. This makes it a local
excitation which is not very sensitive to the length of the lattice or to the
boundary conditions. To confirm this picture, in Fig. 3 we plot $<S_{z}^{f}>$ versus U
for the excited ($S=1, I=0,  S_{z}=1$) state. (Here $S_{z}^{f}$ refers to
the z-component of
the total f-spin of the chain.)
For $U=0$, $<S_{z}^{f}>$ is about 80\% of the total spin and it gets very
close to 100\% for larger U.

In the asymmetric case, 
the gaps tend to a finite value
\cite{Jullien} as $U \rightarrow \infty$.
The fact that the gaps are finite is not
a finite size effect, but rather can be understood in the following
way. For $U=\infty$, states with double f-occupancy are suppressed,
but the proximity of the f-level to the Fermi energy allows the system to
fluctuate between states with no f-electrons in the
local orbital and states with one f-electron.
This leads to hybridization and the formation of the gap.
Another way to see that there is hybridization is to
recall from eq. $(~\ref{eq:Jeff})$ that $J_{eff}=-2V^{2}/\epsilon_{f}$
for $U=\infty$. The presence of a finite $J_{eff}$ leads to hybridization
and gap formation.

{\it Spin Gap.}
For the asymmetric case, we can study how the spin gap
approaches its $U=\infty$ value. We begin by
extrapolating the spin gap to its value at $L \rightarrow \infty$
in the following way:
for each value of U, we plot $\Delta_{s}$ versus $1/L^{2}$,
and obtain a straight line (Fig. 4).
The intercept is then the value for $L \rightarrow \infty$.
To understand why the spin gap is
proportional to $1/L^{2}$, note that the spin density
versus site for the $(S=1,\, S_{z}=1)$ state
is shaped like
a particle-in-a-box wave function of wavelength $\sim L$. This is shown in the
inset of Fig.  4. We then interpret the
spin gap as the kinetic energy of the quasiparticle which is proportional
to the square of the wavevector and goes as $1/L^{2}$.
 
We study the behavior of the spin gap for $L \rightarrow \infty$ and we find that,
for large U with $\epsilon_{f}=0$ in the asymmetric case, it can be fitted
by the form
\begin{equation}
\Delta_{s}(U)=\Delta_{s}(U=\infty) + \frac{Const.}{U^{\alpha}}
\end{equation}
with $\alpha \rightarrow 1$ as $U \rightarrow \infty$.
For $\epsilon_{f} < 0$,
a study of the large U behavior of the gap for short chains
suggests that the spin gap approaches a finite value
as $(\epsilon_{f}+U) ^{-1}$ for the asymmetric case.

\subsection{RKKY Interactions}
We study RKKY interactions between f-spins in the ground state
by calculating the f-spin-f-spin correlation function and the
staggered susceptibility. We first turn our attention to
the correlation function.
\subsubsection{f-spin-f-spin Correlations}

We study the f-spin-f-spin correlations in the ground state due to 
RKKY interactions. In Fig. 5 we show the spin-spin correlation function of the
f-electrons as a function of spatial separation $r$ (measured in units of 
the lattice constant) for different values of U. Fig. 5a corresponds to
the symmetric case and Fig. 5b to the asymmetric case. 

{\it Mixed Valence Case.}
In the mixed valence case, the RKKY correlations are highly suppressed
because the f-orbital is not always occupied.
This can be seen in the f-spin-f-spin correlation function shown in Fig. 5b.
In contrast to the symmetric case, we see that the increase of U by an
order of magnitude does not change the rapid
decrease of the correlations with distance.
 
It is tempting to speculate that another reason why the RKKY correlations
are suppressed is that the effective Kondo coupling $J_{eff}$ becomes very
large as $\epsilon_{f}$ approaches the Fermi level. (Recall that
$J_{eff}=-2V^{2}/\epsilon_{f}$ for $U=\infty$.) In the Kondo lattice model,
RKKY correlations decrease with increasing $J_{eff}$. The problem
with this explanation is that the Schrieffer-Wolff transformation
is no longer valid when $|\epsilon_{f}|\ll \Gamma$ where
$\Gamma=\pi V^{2}\rho$. Even with
more sophisticated techniques \cite{wilkins}, there is no known
analytic expression for $J_{eff}$ when $|\epsilon_{f}|\ll \Gamma$
and $\Gamma/U \ll 1$.

{\it Symmetric Case.}
For the symmetric case we see that as we increase U from 2 to 20, the amplitude of the
oscillations becomes much larger and persists for
longer distances. There are several ways to understand
why RKKY interactions increase with large $U$. First one notes 
that as $U$ increases, the effective Kondo coupling constant $J_{eff}=8V^{2}/U$
decreases relative to the hopping matrix element $t$. Increased hopping
means increased correlations between sites. The decrease of
the gaps also enhances RKKY interactions.
Another way to understand the growth
of RKKY interactions is to note that the Kondo temperature
$T_{K}\sim \exp(-1/\rho J_{eff})$
decreases with increasing $U$ and, as a result, the Kondo compensation cloud increases
in size. Uncompensated f-spins within the cloud develop correlations
\cite{Fye}. A third way to view this was suggested by
Varma and Doniach \cite{vardon}. In the Kondo model they have argued that
RKKY interactions will dominate as $J_{eff}$ decreases
because the RKKY energy scale goes as $J_{eff}^{2}$ while the
Kondo energy scale depends exponentially on $J_{eff}$. 
Opposing this enhancement
of RKKY interactions is the fact that the effective RKKY coupling
will decrease as $J_{eff}$ decreases, but only as $J_{eff}^{2}$.

We find that the amplitude of the RKKY oscillations decreases exponentially
with distance, and it can be fitted by $\exp(-r/\xi)$, where $\xi $
is the correlation length. 
In Fig. 6, we show the behavior of the 
correlation length for the symmetric case as a function of
the effective Kondo coupling constant $J_{eff}=8V^{2}/U$. 
We plot two different sets of parameters and we see that the curves lie on top
of each other, confirming the fact that the relevant parameter in the strong
coupling regime is $J_{eff}$. We also show the results for the one dimensional
Kondo chain, which agree very well for small $J_{eff}$. 
As $J_{eff}$ decreases (i.e., as U increases), 
the correlation length goes up to a maximum and then starts going down. 
There are two ways to interpret this. One view says that
this is to be expected 
since the f-electrons decouple from the conduction electrons 
as $J_{eff}$ goes to zero,
and the correlations between different sites decreases. 
On the other hand, one can interpret the exponential decay of the
spin correlations with distance as a consequence
of the existence of a gap in the excitation spectrum. 
The gap suppresses low energy (long wavelength) magnetic excitations, 
and this greatly reduces long range correlations.
As the gap decreases, one would expect the correlation length to 
monotonically increase roughly as
$\xi \sim v_{Fermi}/\Delta_{S}$ \cite{millislee}. Since
we find a maximum in $\xi$, our data
supports the first point of view, although the 
error bars on $\xi$ are bigger for small $J_{eff}$
because in this regime the effects of the ends affect the wavefunction of
the entire lattice. This can be seen, for example, in the oscillations
of the hybridization energy from site to site. When $J_{eff}$ is small,
these oscillations remain sizable even in the center of the lattice. For
large $J_{eff}$, these oscillations decrease rapidly as one goes away
from the ends of the chain.

\subsubsection{Staggered susceptibility}

In order to further study the magnetic properties of the system, we calculate
the staggered susceptibility $\chi(q)$ 
as a function of the momentum q. To calculate $\chi(q)$, 
we apply a small magnetic field $h=h_{o}\cos(qr)$ in the z direction for different
values of $h_{o}$ and plot $S(q)$ versus $h_{o}$. $S(q)$ is the
Fourier transform of $<0|S_{z}(i)|0>$. (The field couples
to both the f-spins and the conduction spins.) If $h_{o}$ is small enough,
this plot is a straight line whose 
slope is the susceptibility. For this application, $q$ 
has to be a good quantum number. Therefore we use periodic
boundary conditions. This greatly reduces the accuracy of the method,
and we can only study short chains (4 or 6 sites). 

In Fig. 7 we plot 
the staggered susceptibility $\chi(qa=2k_{f}a=\pi)$ versus $U$. We see that in 
the symmetric
case the staggered susceptibility diverges as $U \rightarrow \infty$. 
This is consistent with the growth of RKKY oscillations. In addition, 
as $U \rightarrow \infty$, the f-electrons 
decouple from the conduction electrons, and 
become polarized in an arbitrarily small magnetic field. 
As a result, the susceptibility will diverge for all $q$.
In contrast, in the mixed valence case, as
$U \rightarrow \infty$, the staggered susceptibility $\chi(q=2k_f)$
approaches a finite value $\chi(U=\infty)$. 
$\chi(U)$ does not change much with $U$ because 
of the reduction of the magnetic correlations in the mixed valence regime. 

In the inset of Fig. 7, we show $\chi(q)$ versus $q$ for 
both the symmetric and asymmetric 
cases. In the asymmetric case the peak at $q=2k_{f}$ is greatly reduced, that is, 
$\chi(q)$ shows little structure. This is consistent with neutron
scattering results on CeNiSn $\cite{Mason}$ 
in which they find that $\chi(q)$ is independent of $q$ within experimental error.
In fact the fluctuations we find in $\chi(q)$ are an order of magnitude smaller
than those seen experimentally, though one should keep in mind that our lattices
are small and the values we chose for the parameters 
are not necessarily appropriate for CeNiSn.

In the mixed valence regime the occupation of the f-level is less than
1 per site, which implies that the conduction electron occupation is
greater than 1 per site. One might worry that this implies that the
RKKY wavevector $2k_{f}$ is no longer equal to $\pi$, but rather is equal
to a wavevector that is incommensurate with the lattice. Could this explain
the lack of divergence for $\chi(qa=\pi)$ seen in Fig. 7? The answer is
no. The suppression of RKKY correlations can clearly be seen in Fig. 5b.
Since Fig. 5b is in real space, any RKKY correlations, even those with
an incommensurate wavevector, would show up in this plot.

\subsection{Occupancy of the f-level}

In the symmetric case, the occupation of the f-level is 1 in the ground state
due to particle-hole symmetry. However, in the asymmetric case
we expect the f-level occupation $n_{f}$ to be less than 1 in the mixed valence regime.
We can calculate the f-level occupation $n_{f}$ per site by taking the 
expectation value of $n_{fi}$ on each site
with respect to the ground state wavefunction
and then averaging over the sites. Thus $n_{f}$ is given by 
\be
n_{f}=\frac{1}{L}\sum_{i\sigma}<0|f^{\dagger}_{i\sigma}f_{i\sigma}|0>
\label{eq:nfa}
\ee
where $|0>$ is the ground state.
In Fig. 8, we plot $n_{f}(L=\infty)$ versus U. 
We can extrapolate to infinite lengths because
$n_{f}$ versus $1/L$ follows a straight line 
due to the factor of $1/L$ in eq. (~\ref{eq:nfa}) \cite{nf}.
We see that as $U$ increases, $n_{f}$
decreases from 1 towards a value close
to 0.7, indicating a mixed valence state at large $U$.
For large U, $n_{f}$ follows power law behavior:
\begin{equation}
n_{f}(U)=n_{f}(U=\infty) + \frac{Const.}{U^{\beta}} \label{eq:nfinf}
\end{equation}
where $n_{f}(U=\infty)=0.675\pm 0.001$ and $\beta$ goes to 
$1.4\pm 0.05$ as $U \rightarrow \infty$. 

The results shown in Fig. 8 are for $\epsilon_{f}=0$. We have
studied short chains to see what happens as $\epsilon_{f}$ drops
below zero.  In the mixed valence regime, our study 
suggests that $n_{f}$ has the same form as (~\ref{eq:nfinf}) for large U. 
As $\epsilon_{f}$ becomes more negative, 
$\beta$ increases and approaches a value close to 2 for large $|\epsilon_{f}|$. 
We show in Appendix A that
perturbation theory 
agrees with this result for $V^{2}/|\epsilon_{f}| \ll t$.

\subsection{$U=\infty$}

Our renormalization group approach allows us to study the $U=\infty$ case
by eliminating the doubly occupied f-states from the Hilbert space.
This allows us to compare our results
to the predictions of approximation schemes which are applied
in this limit, e.g., the slave 
boson approach $\cite{Riseborough}$
and the Gutzwiller approximation $\cite{Rice}$. 

We study what happens as we reduce $V$ in the cases when
$\epsilon_{f}=0$ and $\epsilon_{f}=-0.5$.
In Fig. 9 we compare the spin gap with the one predicted by the slave boson 
mean field technique $\cite{Riseborough}$. For the case $\epsilon_{f}=0$
shown in Fig. 9a, there is overall qualitative agreement, although the
slave boson mean field theory overestimates the size of the 
gap. It is important to 
point out that the slave boson curve has been calculated using a 
constant density of states while the numerical results use a tight
binding density of states. Nevertheless, when $\epsilon_{f}=0$,
both give power law behavior for small V \cite{Jullien}: $\Delta_{s} \sim V^{4}$. 
For the case $\epsilon_{f}=-0.5$, the situation is quite different as can
be seen in Fig. 9b. 
There is qualitative agreement for $V$ greater than 0.5 but then the slave boson 
curve drops very rapidly for small $V$. 
A similar situation can be seen in Figs. 10a and 10b for $1-n_{f}$.

To understand these results let us take a closer look at the slave 
boson mean field solution $\cite{Riseborough}$. The number of slave
bosons $\sum_{i} b^{\dagger}_{i}b_{i}$ represents the number of sites with no f-electrons.
Each site is subject to the constraint 
\be
b^{\dagger}_{i}b_{i} + \sum_{\sigma}f^{\dagger}_{i,\sigma}f_{i,\sigma}=1 \label{slave}
\ee
Since the slave boson number operator is positive definite, each site is
constrained to have less than one f-electron.
In the mean field approximation the boson operator is 
approximated by its mean value $a_{o}$, i.e., $b^{\dagger}_{i}=b_{i}=a_{o}$, where
$a_{o}$ is a real number. This amounts to suppressing double occupancy
of the f-level only on average.  Using slave bosons, an effective Hamiltonian can be
written in which doubly occupied sites have been projected out. As a result,
there is no U-term. 
The resulting effective Hamiltonian is then equivalent to the $U=0$ Anderson
Hamiltonian (~\ref{eq:Hamiltonian})
with an effective hybridization
\begin{equation}
V_{eff}^{2}=V^{2} a_{o}^{2} =V^{2}(1-n_{f})   \label{vefsb}
\end{equation}
The slave boson technique recovers the expected
picture of two hybridized bands with a gap.
The expression for the gap is the same as for the $U=0$ Anderson Hamiltonian
with this effective hybridization. For small $V$, the gap goes as $V_{eff}^{2}/t$.
The f-level is also renormalized to $\tilde{\epsilon_{f}}$, and self-consistent
equations for both $\tilde{\epsilon_{f}}$ and $a_{o}$ can be obtained 
$\cite{Riseborough}$. 
The small V expansion of the self-consistent equations gives
\begin{eqnarray}
\nonumber \tilde{\epsilon_{f}} & = &2ta_{o}^{2} \\
a_{o}^{2} & = & \frac{4t^{2}}{V^{2}}  
\exp(\frac{2t(\epsilon_{f}-\tilde{\epsilon_{f}})}{V^{2}}) \label{sb}
\end{eqnarray}

When $\epsilon_{f}<0$ and V is small enough such that $a_{o}$ is very
small, then $\tilde{\epsilon_{f}}$ in the exponential can be neglected, and
$a_{o}$ goes to zero as $\exp(-2t|\epsilon_{f}|/V^{2})$.
($\tilde{\epsilon_{f}}>0$.) Thus the gap will depend exponentially on $V$.
When $\tilde{\epsilon_{f}}$ becomes greater than $\epsilon_{f}$,
$a_{o}^{2}=1-n_{f}$ crosses over from exponential to power law behavior.
If $|\epsilon_{f}| \ll \tilde{\epsilon_{f}}$, then $\epsilon_{f}$
in the exponent of equation (~\ref{sb}) can be neglected. Plugging
in $\tilde{\epsilon_{f}}=2ta_{o}^{2}$ gives a transcendental equation for
$a_{o}$. Solving this numerically, we find that $a_{o}^{2}\sim V^{2}$.
As a consequence, the gap goes as $V^{4}$.
Thus the slave boson mean field approach predicts that the gap
crosses over from exponential dependence on $V$ to power law dependence
as $\epsilon_{f}$ approaches 0 from below.
However, our numerical renormalization group 
results do not show such a change in behavior. 
Although it might be argued that finite size systems will give power law rather than
exponential behavior, increasing the length of the chain
does not indicate any evidence
of such a dramatic crossover. In addition, perturbation theory 
for an infinite chain
shows that $1-n_{f}$ goes as $V^{2}$ when $V^{2}/\epsilon_{f} \ll t$
(see Appendix A). 
This is exactly what we find for $\epsilon_{f}=-0.5$ with small $V$. 

The reason for this disagreement between the slave boson
calculation and the numerical renormalization group approach
may be that the mean field approximation
neglects spin fluctuations. 
Spin fluctuations are crucial in the Kondo regime.
They also give rise to RKKY interactions.

The $U=\infty$ limit has also been treated by the Gutzwiller approximation
\cite{Rice}. 
It predicts a ferromagnetic ground state when the system goes towards the Kondo 
regime where $V^{2}/\epsilon_{f}t\ll 1$. However, Rice and Ueda
\cite{Rice} only considered uniform magnetic states, as opposed to antiferromagnetic
states which are favored by RKKY interactions.
Using a saddle point formulation
of the Gutzwiller method, Reynolds {\it et al.} \cite{Reynolds} suggest that
the Gutzwiller approximation is biased too much in favor of a 
ferromagnetic state.
However, they still find a ferromagnetic
instability in the Kondo regime. We do not see any 
evidence of ferromagnetism in any of 
the cases we studied. We always find
an $S=0$ ground state. We find that strong antiferromagnetic 
correlations appear as the system goes 
towards the Kondo regime.
\section{Conclusions}
We have studied the one dimensional Anderson
lattice at half filling using the density matrix 
numerical renormalization group technique. 
We considered the symmetric Anderson model with $\epsilon_{f}=-U/2$, as well
as the asymmetric case in which we set
$\epsilon_{f}=0$ in order to study the mixed valence regime. 
Since many real materials lack particle-hole symmetry
and are likely to be in the mixed valence regime,
we concentrated on the asymmetric model with $\epsilon_{f}=0$.
Our results suggest that some Kondo insulators behave as mixed valence 
systems, and therefore should be modeled by the Anderson lattice model 
in the mixed valence regime as opposed to the Kondo lattice or
the Anderson lattice in the Kondo regime. 

For all
the values of the parameters that we have used, including $\epsilon_{f}\leq 0$
in the asymmetric case, we find that this is
an insulating system with gaps.
As $U\rightarrow \infty$ in the symmetric case,
the conduction electrons decouple from the f-electrons and the
gaps approach the values that they have for free electrons.
However, in the 
asymmetric case the gaps approach a finite value
in the large $U$ limit because the conduction electrons do not decouple
from the f-electrons. There is hybridization due to charge fluctuations
in and out of the band of local orbitals whose energy is at
or near the Fermi energy. 

In both the symmetric and asymmetric
cases we find that the charge and quasiparticle gaps are larger than the
spin gap for $U>0$. 
However, we found that the relative values are different
in each case. 
For the mixed valence case the ratio between the charge gap
and the spin gap is roughly two in the strong coupling regime (large U). 
This is 
in agreement with reference
\cite{bucher} in which the authors report a ratio between the charge gap 
and the spin gap of $Ce_{3}Bi_{4}Pt_{3}$ equal to 1.8.
In contrast, the strong coupling limit of the symmetric case
(Kondo regime) gives a much larger ratio between the two gaps. 
In fact, it was shown by Nishino and Ueda \cite{Nishino} that 
this ratio diverges as the 
Coulomb interaction U goes to infinity. Our result, together with the fact that
the occupation of the localized orbital is 0.865 at T=0 \cite{Kwei}, 
suggests that $Ce_{3}Bi_{4}Pt_{3}$ behaves like a mixed valence compound at
low temperatures.

In the asymmetric case the nature 
of the lowest lying charge excitation changes as $U$ increases. 
For small $U$ the charge state is made out of two spin-1/2 excitations,
while for larger $U$, it becomes a state consisting of two spin-1 excitations.

The fact that the gaps are smaller for large
$U$ than for $U=0$ is consistent with the fact that band structure
calculations on FeSi \cite{Fu,Mattheiss}, which ignore many body correlations,
predict values for the gap that are larger than the optical gap measured experimentally
\cite{schlesinger}.

We have studied the RKKY interactions by calculating the
f-spin-f-spin correlation function as well as the staggered
susceptibility $\chi(q)$. We find that the amplitude of the RKKY
oscillations decays exponentially with distance. For the symmetric Anderson model, 
the RKKY interactions become important in the strong coupling regime
($8V^{2}/U \ll t$). This can be seen, for example,
in the divergence of the susceptibility $\chi(q=2k_f)$ as
$U\rightarrow \infty$. This stands in sharp contrast to the
asymmetric case in the mixed valence regime where
the RKKY correlations are strongly suppressed due to
the reduction in the f-occupancy.
This agrees with Varma's
argument \cite{Varma}
that magnetic correlations are suppressed in the mixed valence case.
Our findings are also consistent with neutron scattering results
on CeNiSn $\cite{Mason}$ which found $\chi(q)$ independent of $q$.
Further experiments would be necessary to confirm that CeNiSn is a mixed
valence compound. 
One such measurement would be photoemission which can measure the occupation
of the f-level. 
In the mixed valence case that we consider 
(the localized level right at 
the Fermi energy and $\Gamma \sim 1$), we find that $n_{f}$ varies from
1 at $U=0$ to a value close to $0.7$ at $U=\infty$.

We would like to thank Steve White, Herv\'{e} Carruzzo, 
Andy Millis, Luiz Oliveira, Doug Scalapino, George Gruner,
Kazuo Ueda, and John Wilkins for helpful
discussions. This work was supported in part by ONR Grant No. N000014-91-J-1502
and an allocation of computer time from the University of California, Irvine.
C. C. Yu is an Alfred P. Sloan Research Fellow.

\subsection{Appendix A}

In this appendix we use perturbation theory to show that for small $V$, 
$1-n_{f} \, \sim V^{2}$ for $U=\infty$. We also show that
$n_{f}(U)-n_{f}(U=\infty) \, \sim U^{-2}$.
Consider the Anderson Hamiltonian (~\ref{eq:Hamiltonian}) for the case
where the hybridization matrix element is zero, i.e., $V=0$, $\epsilon_{f}<0$, 
and $\epsilon_{f}+U>0$.
In this case the ground state consists of a half-filled system with
one electron per site in the f-band. The f-electrons states are degenerate with 
respect to their spin configurations. 

For $V>0$ but $ V^{2}/2t(\epsilon_{f}+U) \ll 1$ and
$ V^{2}/2t\epsilon_{f} \ll 1 \,$, perturbation theory can be used to
calculate the correction to the ground state energy. The first non-zero
contribution is the second order term:
\begin{equation}
\Delta E_{o}= \sum_{m\neq 0} \frac{|<m|H_{1}|0>|^{2}}{E_{o}-E_{m}}
\end{equation}
where $H_{1}$ is the hybridization term. The only possible states that
contribute to the sum are those that have one hole in the Fermi sea and 
an extra electron in the f-band or vice-versa. Also, the fact that 
$k_{f}a=\pi/2=\pi-k_{f}a$ allows one to write these two contributions under the same sum:
\begin{equation}
\Delta E_{o}=V^{2}\sum_{k>k_{f}} \left[ \frac{1}{\epsilon_{f}-\epsilon_{k}} -
                        \frac{1}{\epsilon_{k}+\epsilon_{f}+U} \right]
\end{equation}
where $\epsilon_{k}$ is the energy of the conduction band. In our case, 
$\epsilon_{k}=-2tcos(ka)$.
For the symmetric case, $\epsilon_{f}=-U/2$, this expression coincides with the one obtained
by Blankenbecler {\it et al.} $\cite{Blankenbecler}$.
The total ground state energy to second order is given by
\begin{equation}
e_{o}=\frac{E_{o}}{N} = \epsilon_{f} + \frac{\sum_{k<k_{f}} \epsilon_{k}}{N}
       + \frac{\Delta E_{o}}{N} 
\end{equation}
The occupation of the f-orbital can be obtained by the relation
\begin{equation}
n_{f}=\frac{\partial e_{o}}{\partial \epsilon_{f}}=1-\frac{V^{2}}{N}
                 \sum_{k>k_{f}} \left[\frac{1}{(\epsilon_{k}+\epsilon_{f}+U)^{2}}
            +   \frac{1}{(\epsilon_{f}-\epsilon_{k})^{2}}\right]
\end{equation}
For $U=\infty$, this expression gives:
\begin{equation}
1-n_{f}(U=\infty)=\frac{V^{2}}{N}
                  \sum_{k>k_{f}} \frac{1}{(\epsilon_{f}-\epsilon_{k})^{2}}
                  \label{eq:1-nf}
\end{equation}
This implies that for small $V$, $1-n_{f} \, \sim V^{2}$. Note that for arbitrary $U$ 
\begin{equation}
n_{f}(U)-n_{f}(U=\infty)=\frac{V^{2}}{N}
             \sum_{k>k_{f}} \frac{1}{(\epsilon_{k}+\epsilon_{f}+U)^{2}}
             \label{eq:nf}
\end{equation}
Thus for large $U$, $n_{f}(U)-n_{f}(U=\infty) \, \sim U^{-2}$.

\newpage
\begin{center}
FIGURE CAPTIONS
\end{center}
\noindent
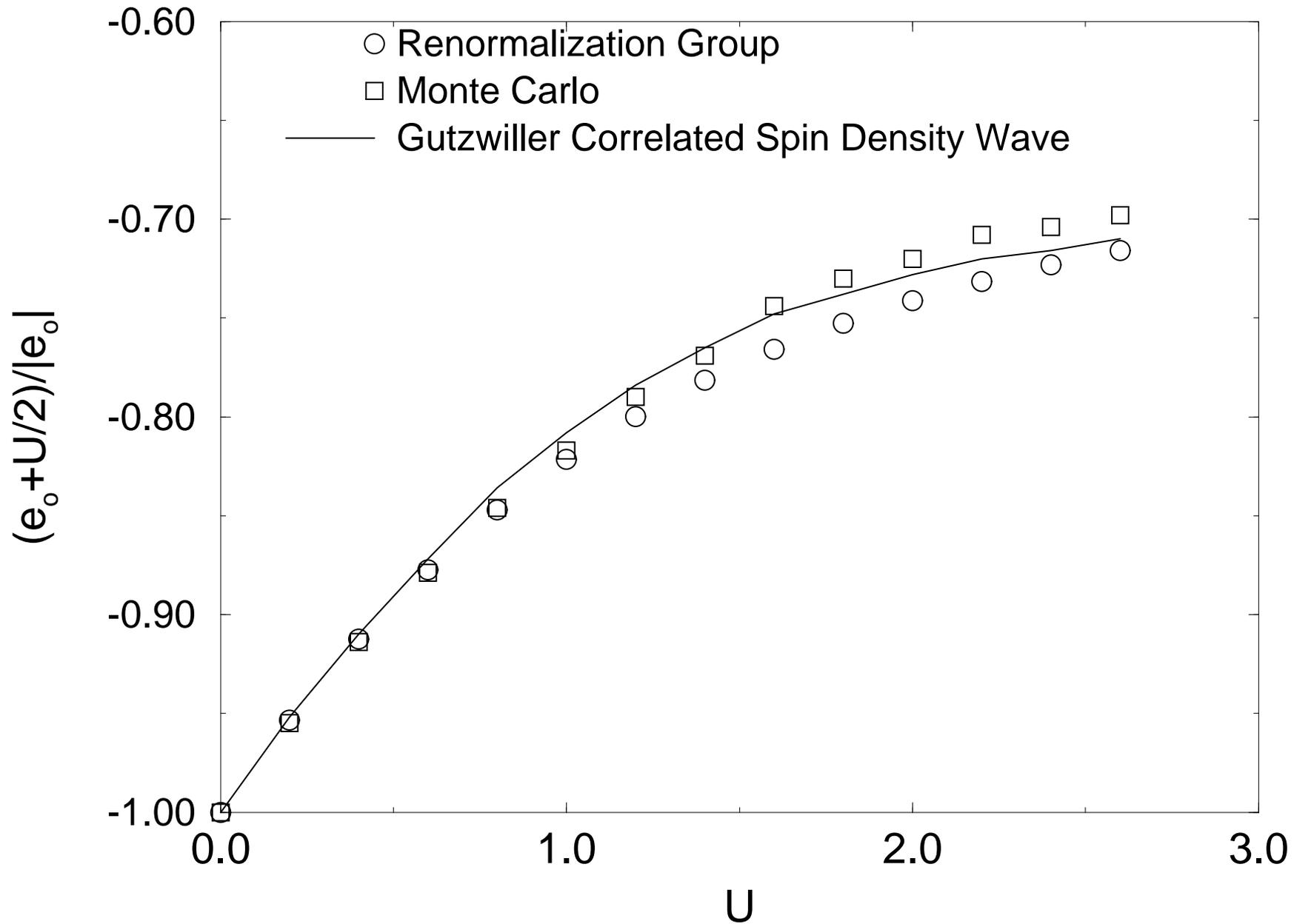
Figure 1. $(e_{o}+U/2)/|e_{o}|$ vs. $U$ for the symmetric Anderson lattice with
V/2t=0.375. $e_{o}$ is the ground state energy per site.
Our results are in good agreement with Gutzwiller \cite{Gulacsi} and
Monte Carlo \cite{Blankenbecler} results. 

\noindent Figure 2a. Gaps vs. U for the symmetric case 
(t=1, V=1, $\epsilon_{f} =-U/2$). Open symbols are for the quasiparticle gap. 
Filled symbols are for the spin gap and the charge gap. Note that
$\Delta_{s}< \Delta_{qp} <\Delta_{c}$ for $U>0$.

\noindent Figure 2b. Gaps vs. U for the asymmetric case 
(t=1, V=1, $\epsilon_{f}$=0). The charge and quasiparticle gap cross at 
$U \sim 2$. Note that the spin gap is smaller than both of them.

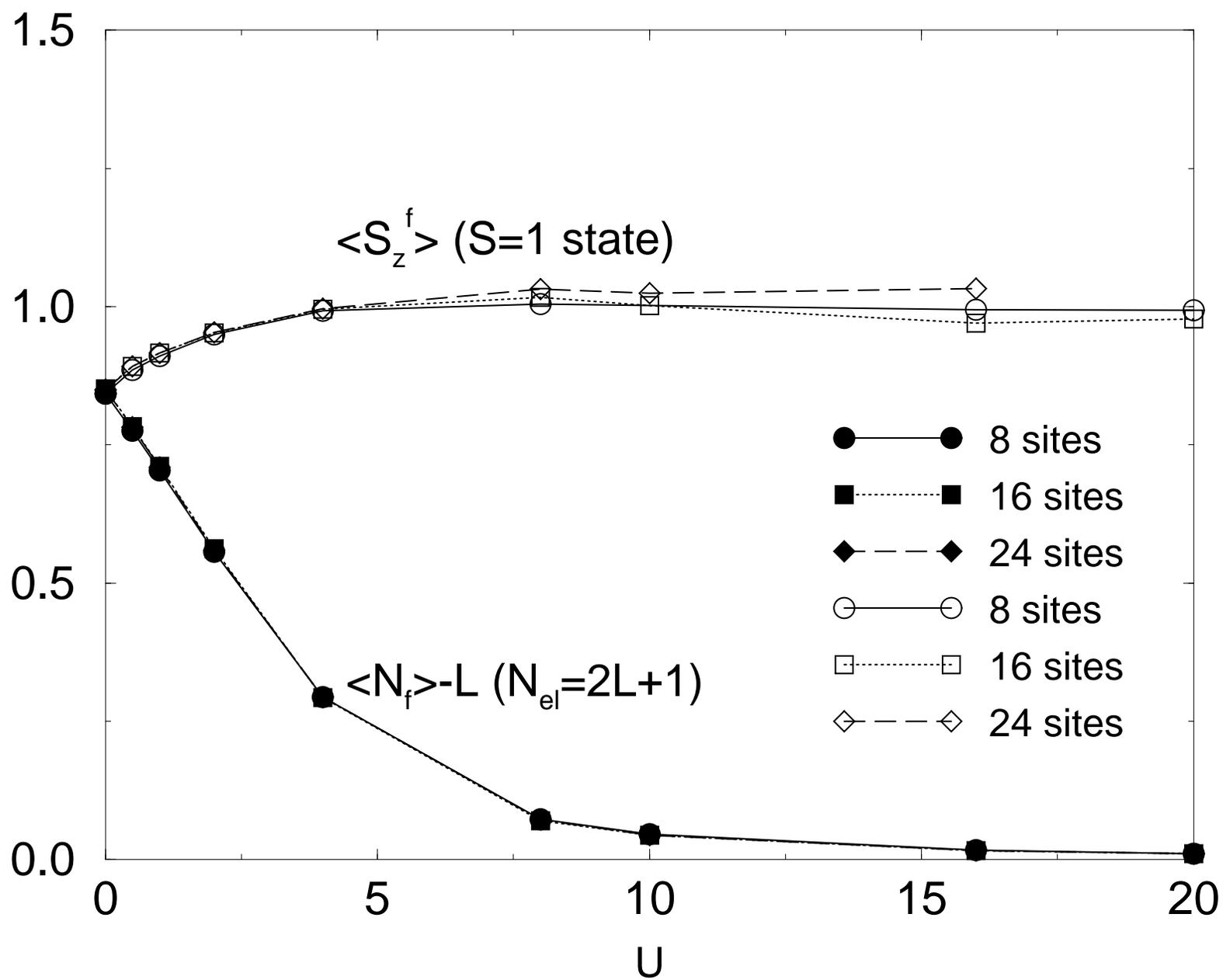
\noindent Figure 3. Character of the spin and quasiparticle excitations
vs. $U$. The open symbols are for $<S_{z}^{f}>$ in the lowest excited
S=1 state at half filling.
The filled symbols are for $<N_{f}>-L$ in the ground state
with $2L+1$ electrons.
The spin excitations have mainly f-character while the particle 
excitations have the character of conduction electrons in 
the strong coupling regime.

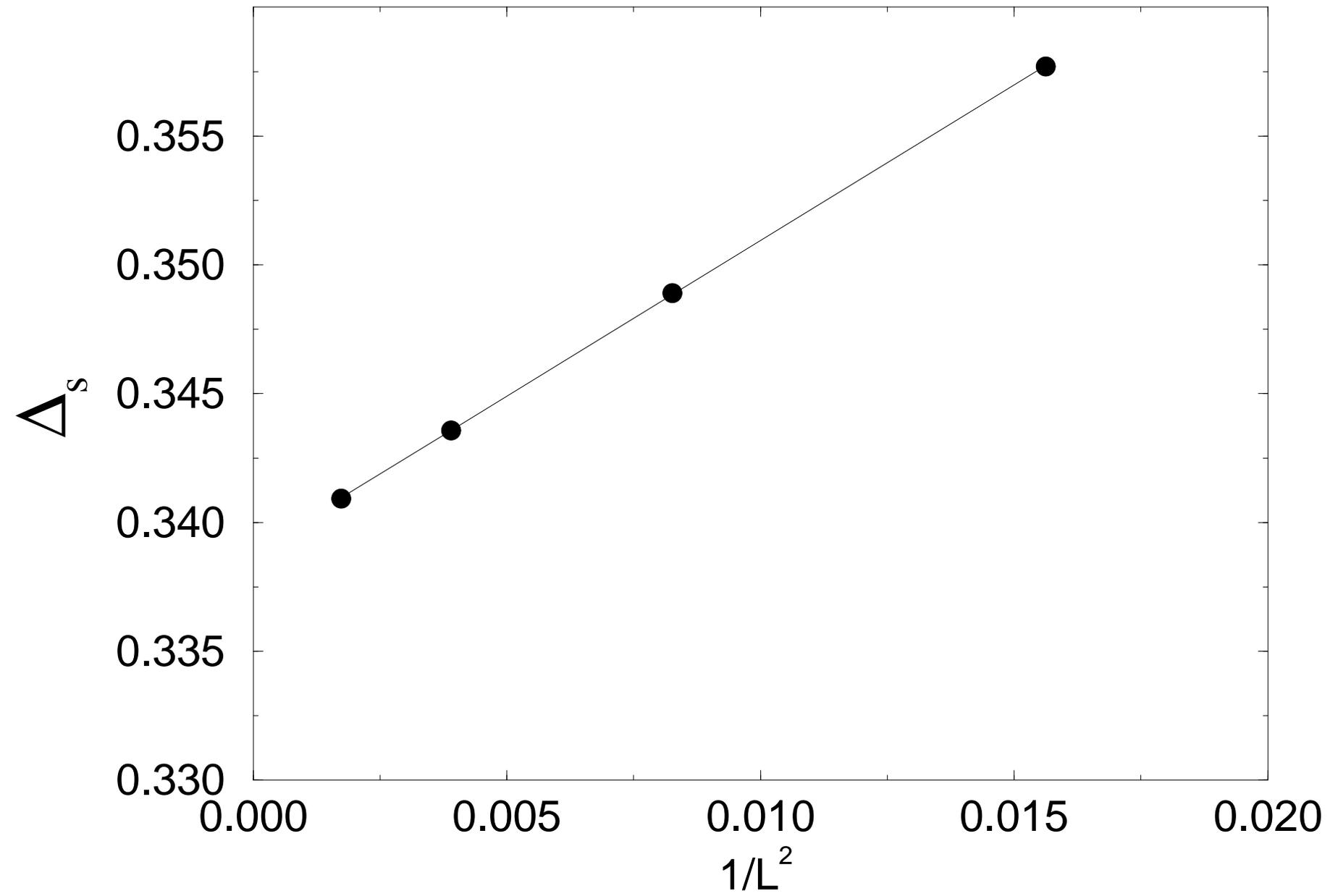
\noindent Figure 4. Spin gap vs. $1/L^{2}$ in the asymmetric case 
          (t=1, V=1, $\epsilon_{f}$=0, U=16). The intercept of the
          linear fitting gives the $L=\infty$ value.
	Inset: $S_{z}$ on a site vs. site for the lowest excited (S=1, $S_{z}=1$) state
	on a 24 site lattice at half filling.

\noindent Figure 5a. f-spin-f-spin correlation function versus distance r apart 
	   for the symmetric case
           ($t=1, \,V=1,\, \epsilon_{f}=-U/2$, L=24). The oscillations increase
           in amplitude and persist for longer distances as U increases from 2
           to 20.

\noindent Figure 5b. f-spin-f-spin correlation function versus distance r apart 
	for the asymmetric case
           ($t=1, \,V=1, \,\epsilon_{f}=0, L=24 $). As U increases from 2 to 20, 
          neither the amplitude of the oscillations nor the correlation 
          length change significantly.

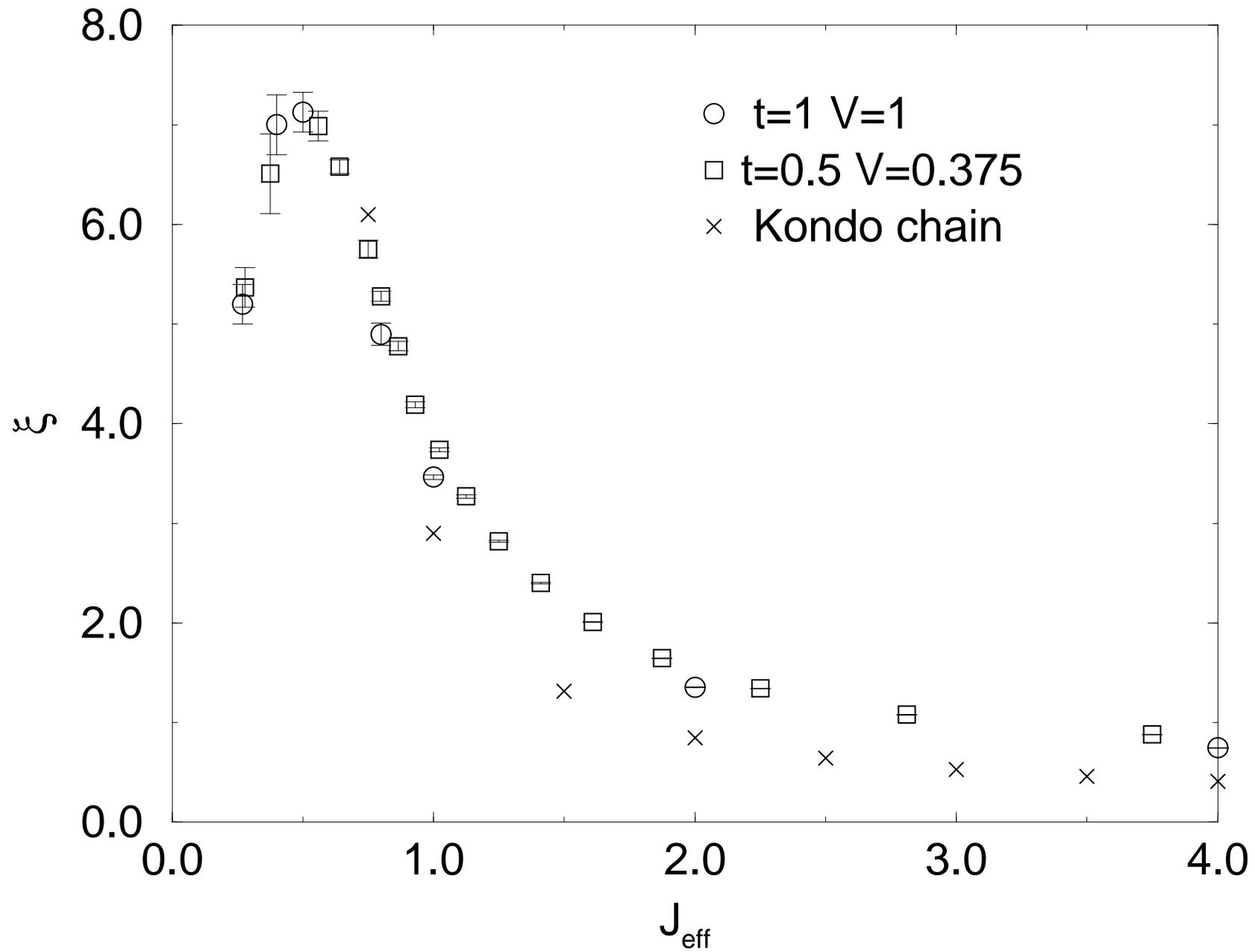
\noindent  Figure 6. Correlation length vs. $J_{eff}=8V^{2}/U$ in the symmetric case.
          The curves for different sets of parameters fall on top of each 
          other for $J_{eff}<1$. The data for the Kondo chain is taken 
          from reference $\cite{Yu1}$. All data is for L=24.

\noindent Figure 7. Staggered susceptibility vs. U for short chains with periodic 
          boundary conditions (t=1, V=1). 
          In the asymmetric case the susceptibility
          shows little change with U due to the 
          suppression of the magnetic correlations in the ground state.
	Inset: Staggered susceptibility vs. q for short chains with $U=16$.
	Notice that the asymmetric case (filled symbols) shows little
	dispersion indicating again that RKKY interactions are suppressed in
	the asymmetric case.

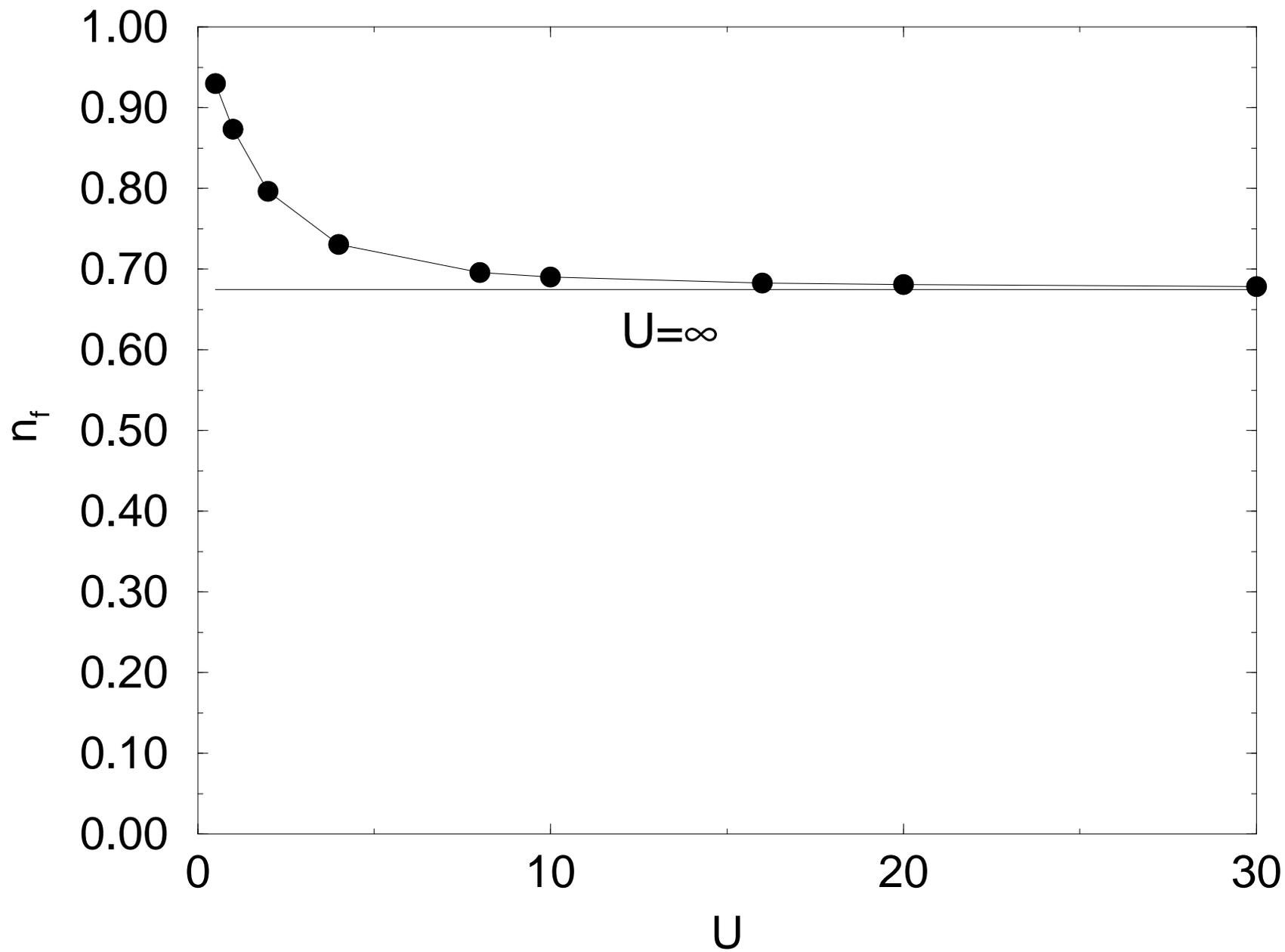
\noindent Figure 8. The occupation $n_{f}$ of the f-level  vs. U for $L=\infty$
          in the asymmetric case (t=1, V=1, $\epsilon_{f}=0$).
          As $U \rightarrow \infty$, $n_{f}$ approaches a value close to 0.7.

\noindent Figure 9a. $\log(\Delta_{s})$ vs. $\log(V)$ for the $U=\infty$ case 
           (t=1, $\epsilon_{f}=0$). The slave boson results 
	   agree qualitatively with our numerical results.

\noindent Figure 9b. $\log(\Delta_{s})$ vs. $\log(V)$ for the $U=\infty$ case 
           (t=1, $\epsilon_{f}=-0.5$). For small V, the 
           slave boson results deviate significantly from our data.

\noindent Figure 10a. $\log(1-n_{f})$ vs. $\log(V)$ for the $U=\infty$ case 
           (t=1, $\epsilon_{f}=0$). The slave boson results 
           agree qualitatively with our numerical results.

\noindent Figure 10b. $\log(1-n_{f})$ vs. $\log(V)$ for the $U=\infty$ case 
           (t=1, $\epsilon_{f}=-0.5$). For small V, the 
           slave boson results deviate significantly from our data.

\begin{thebibliography}{99}
\bibitem{nf} The value of $n_{fi}$ is much closer to $n_{f}(L=\infty)$ at sites
in the center of the lattice than at sites near the ends. From eq.
(~\ref{eq:nfa}), we see that the contribution to $n_{f}$ from sites near the
ends falls off as $1/L$.
\end{thebibliography}
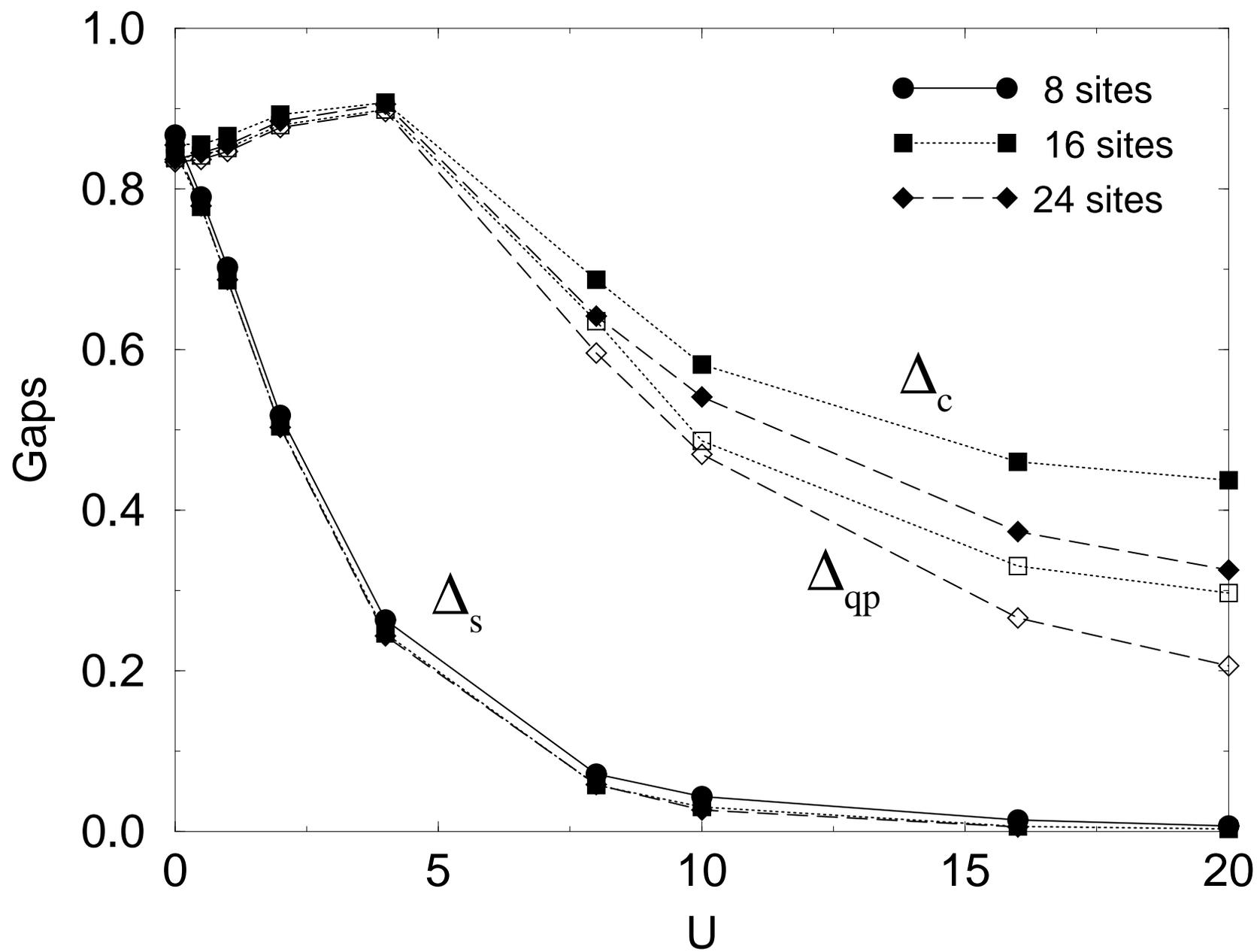
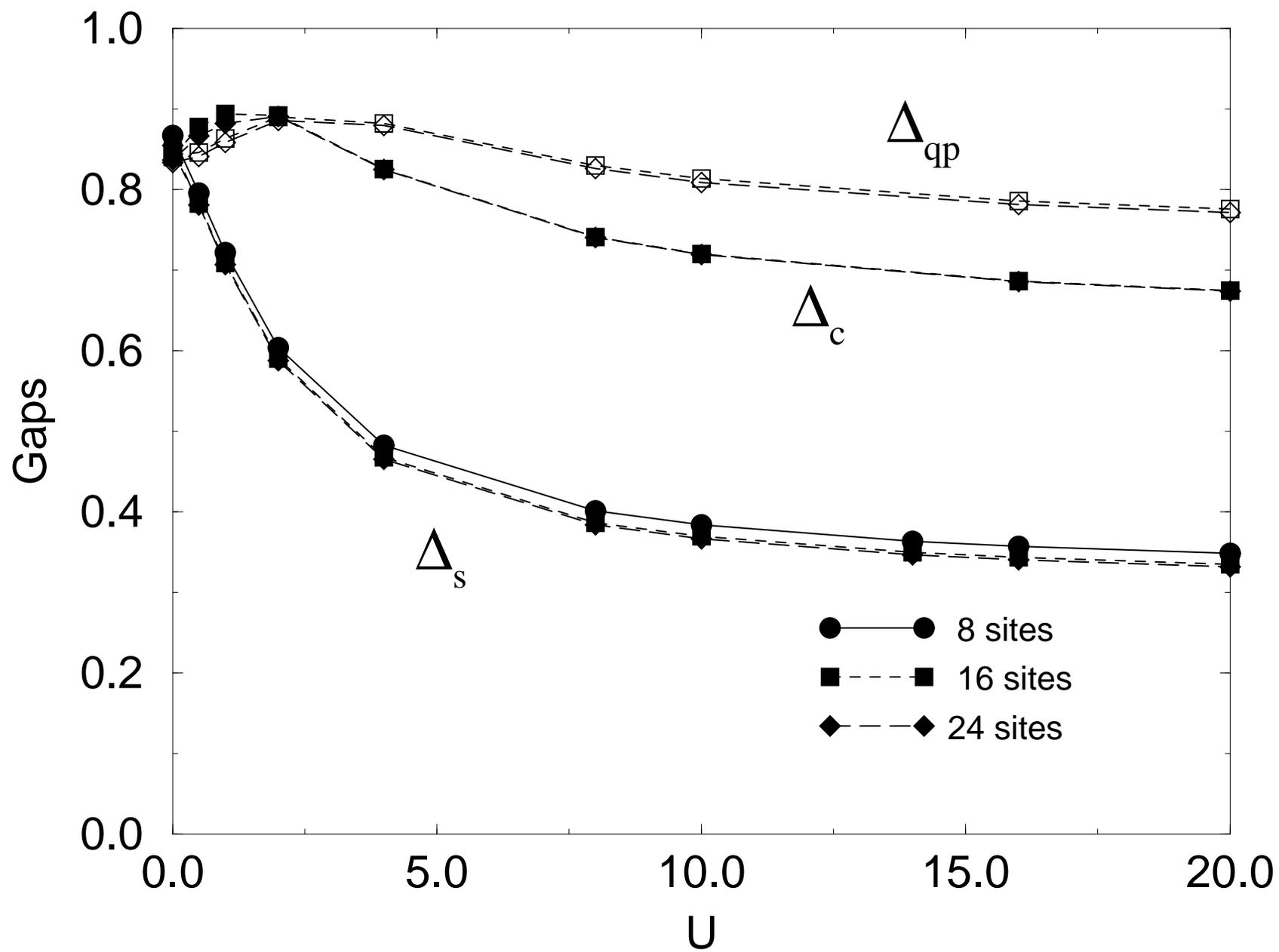
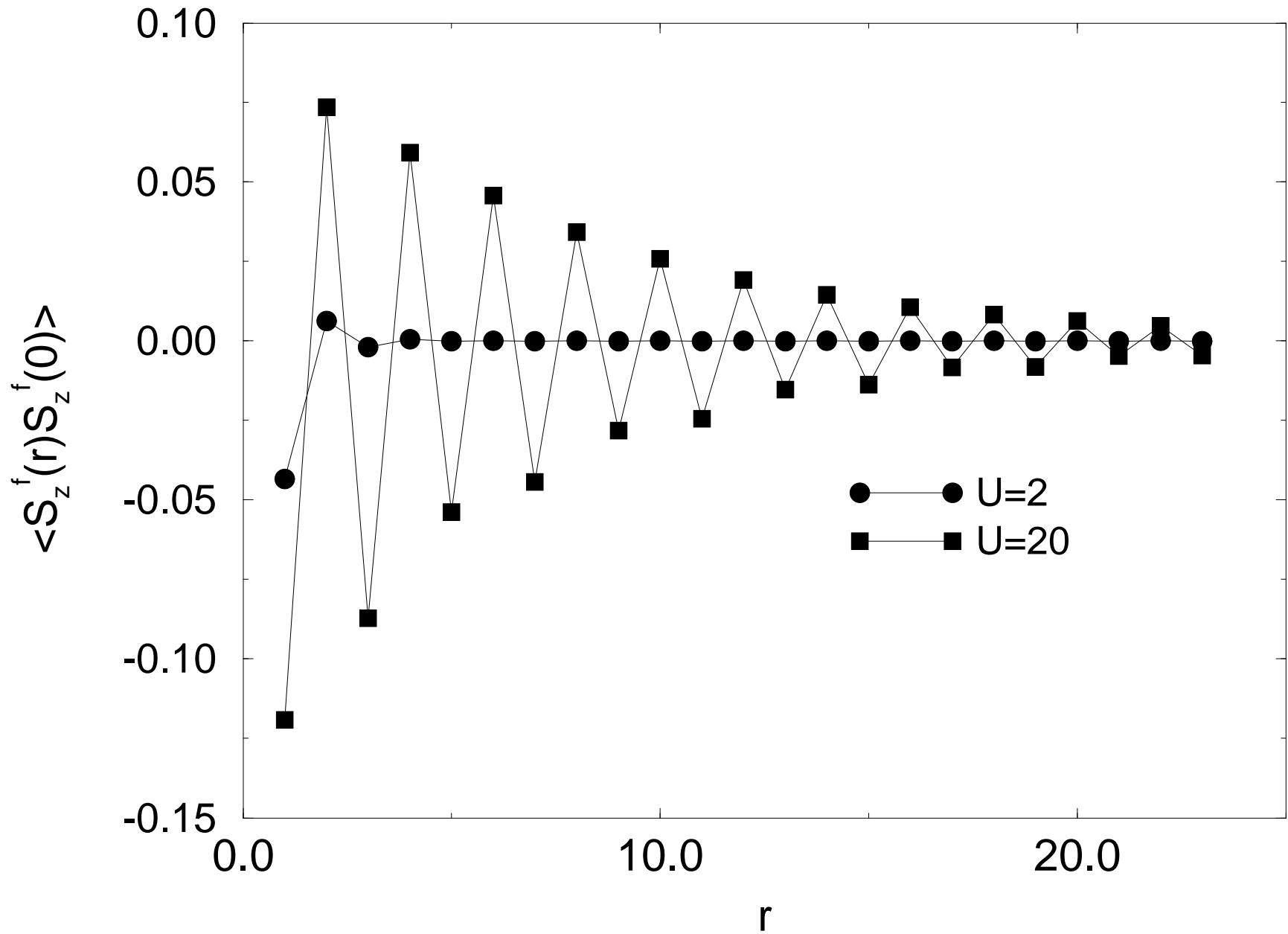
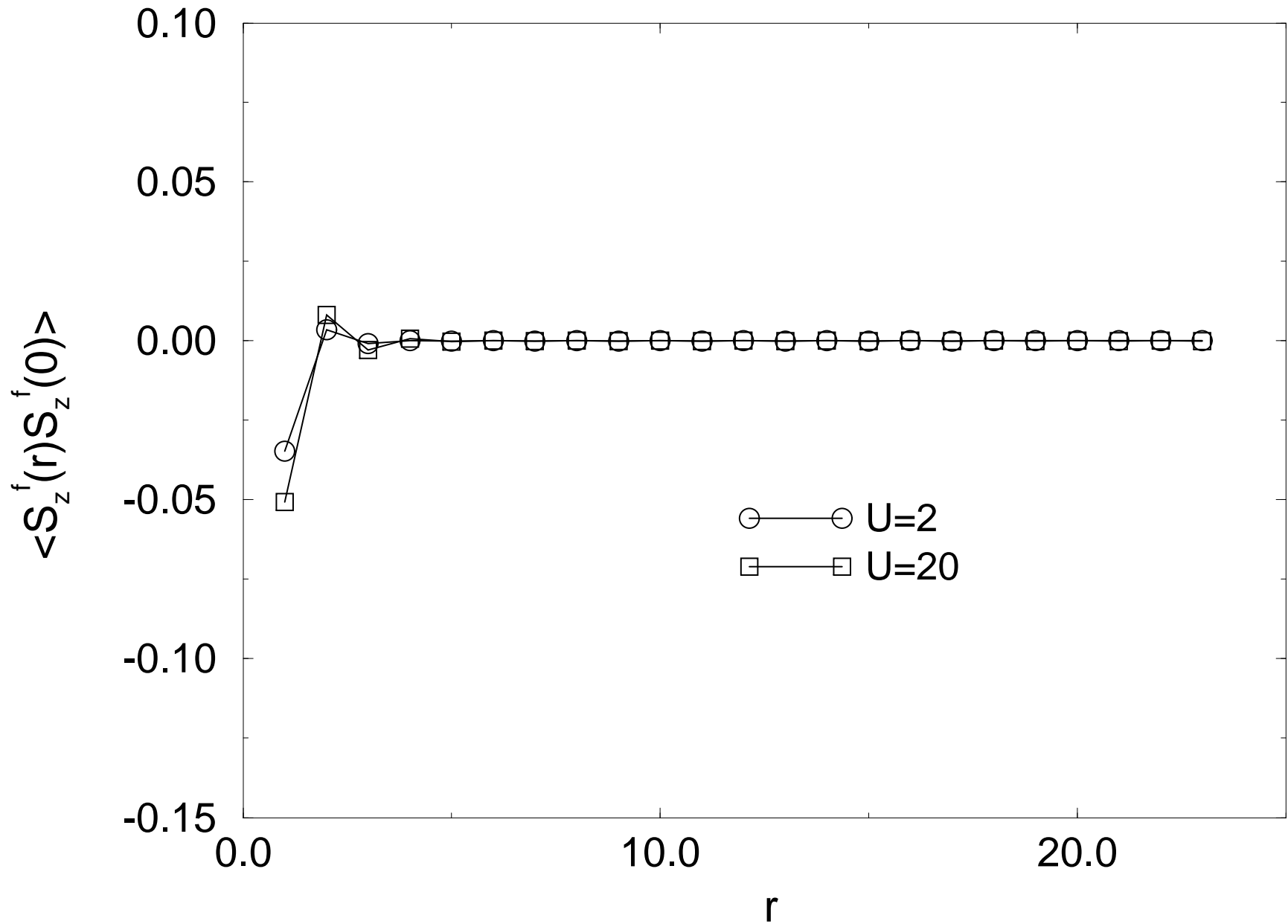
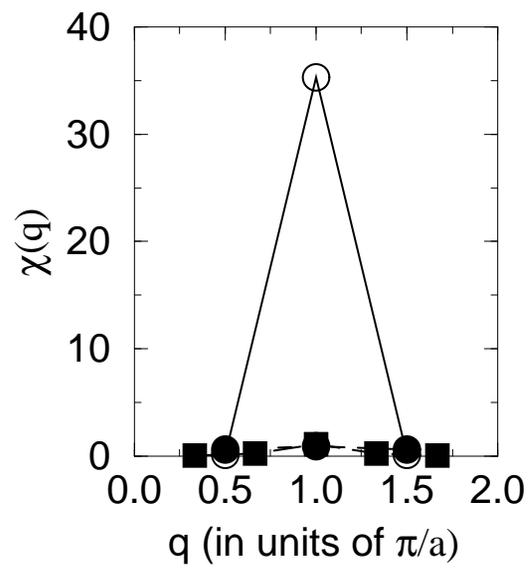
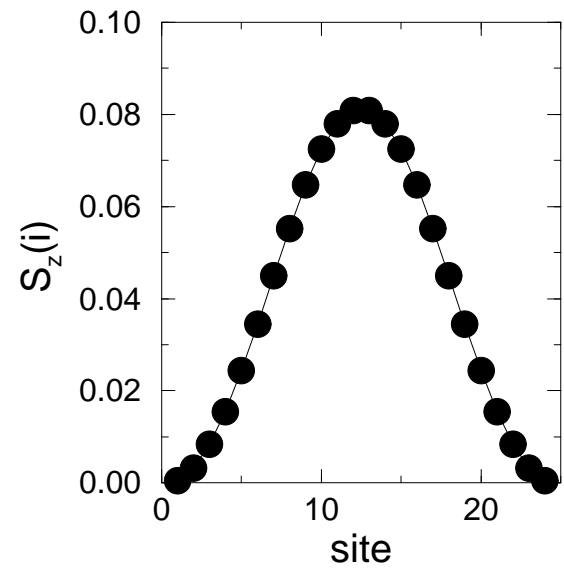
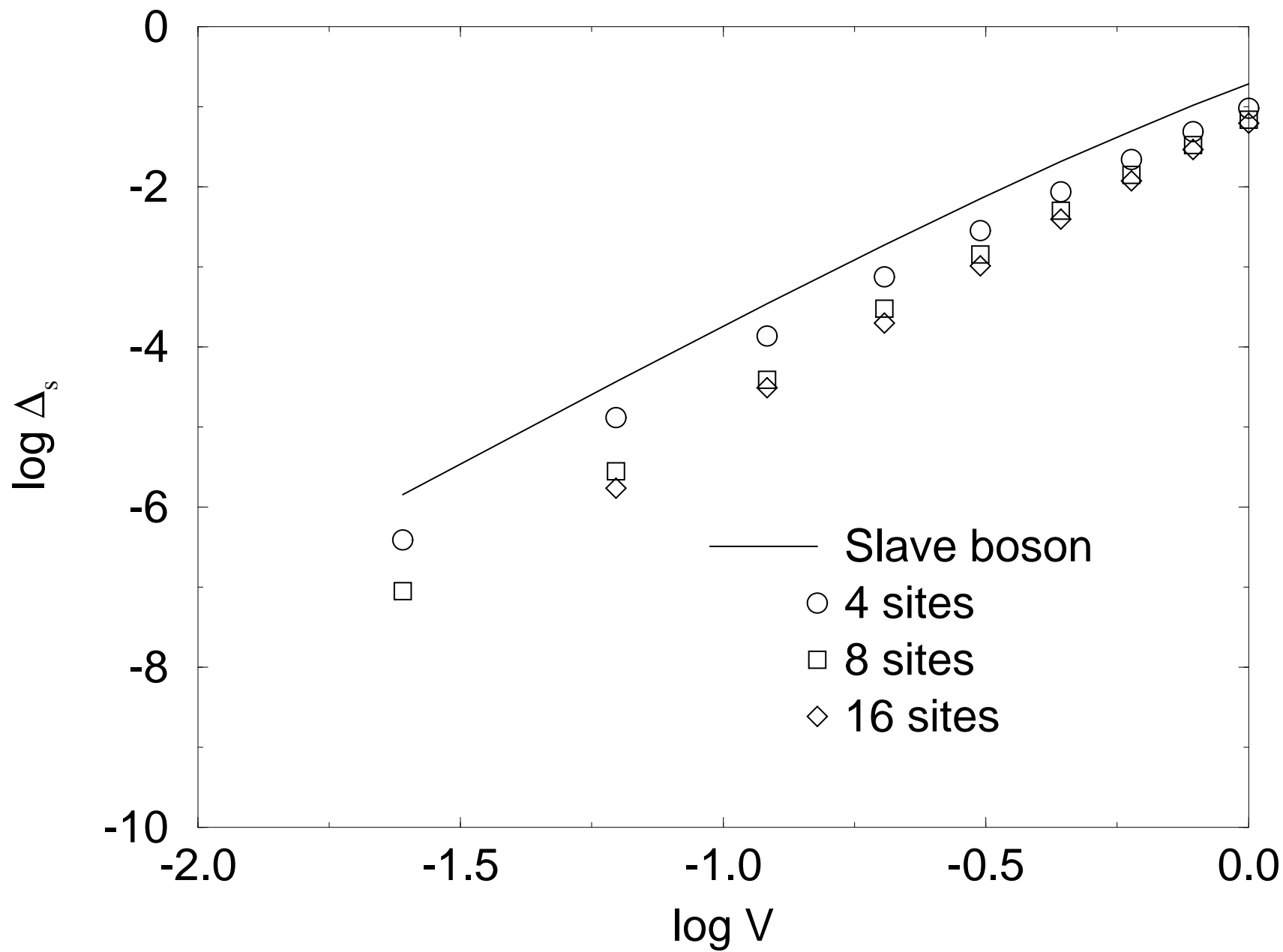
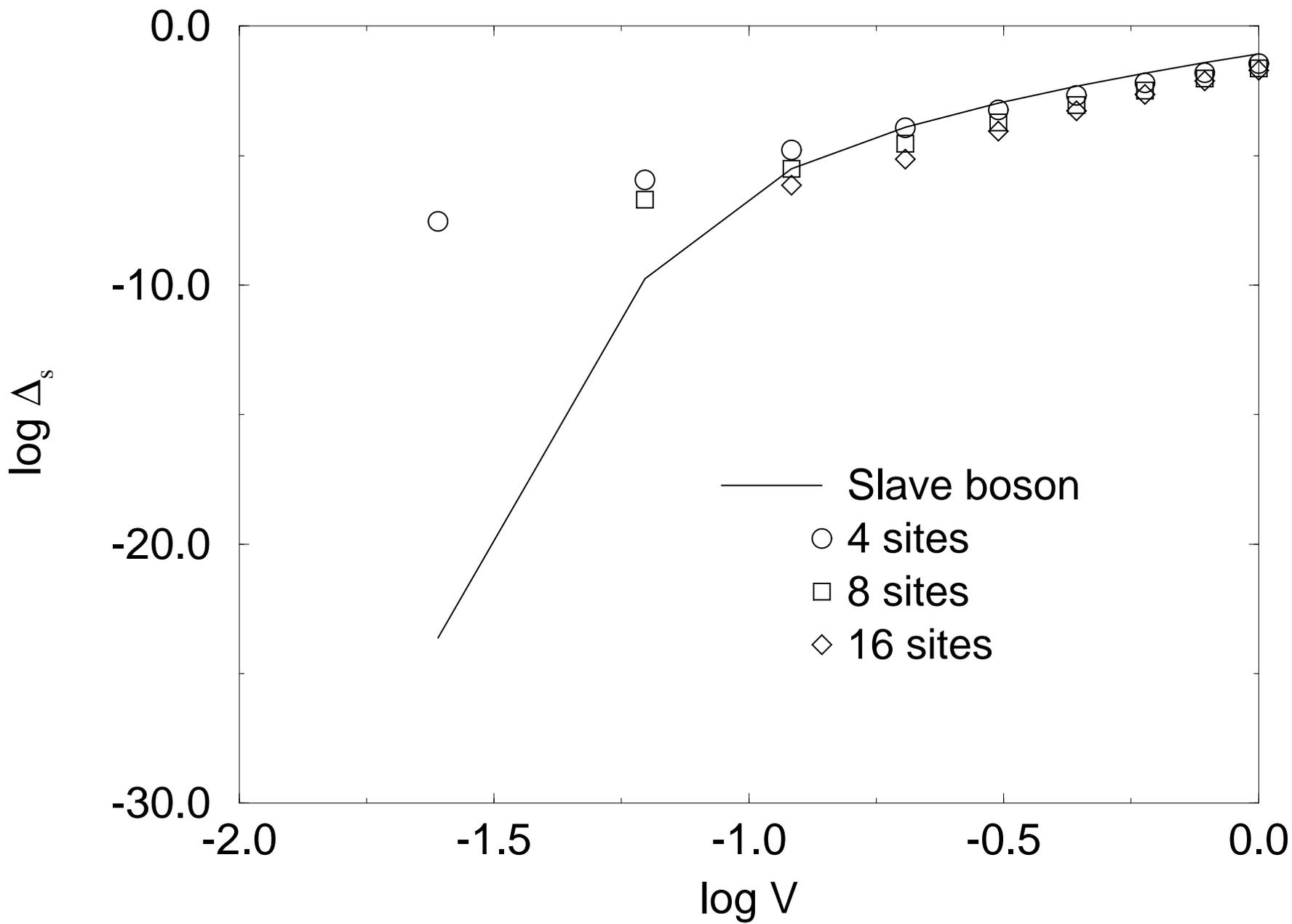
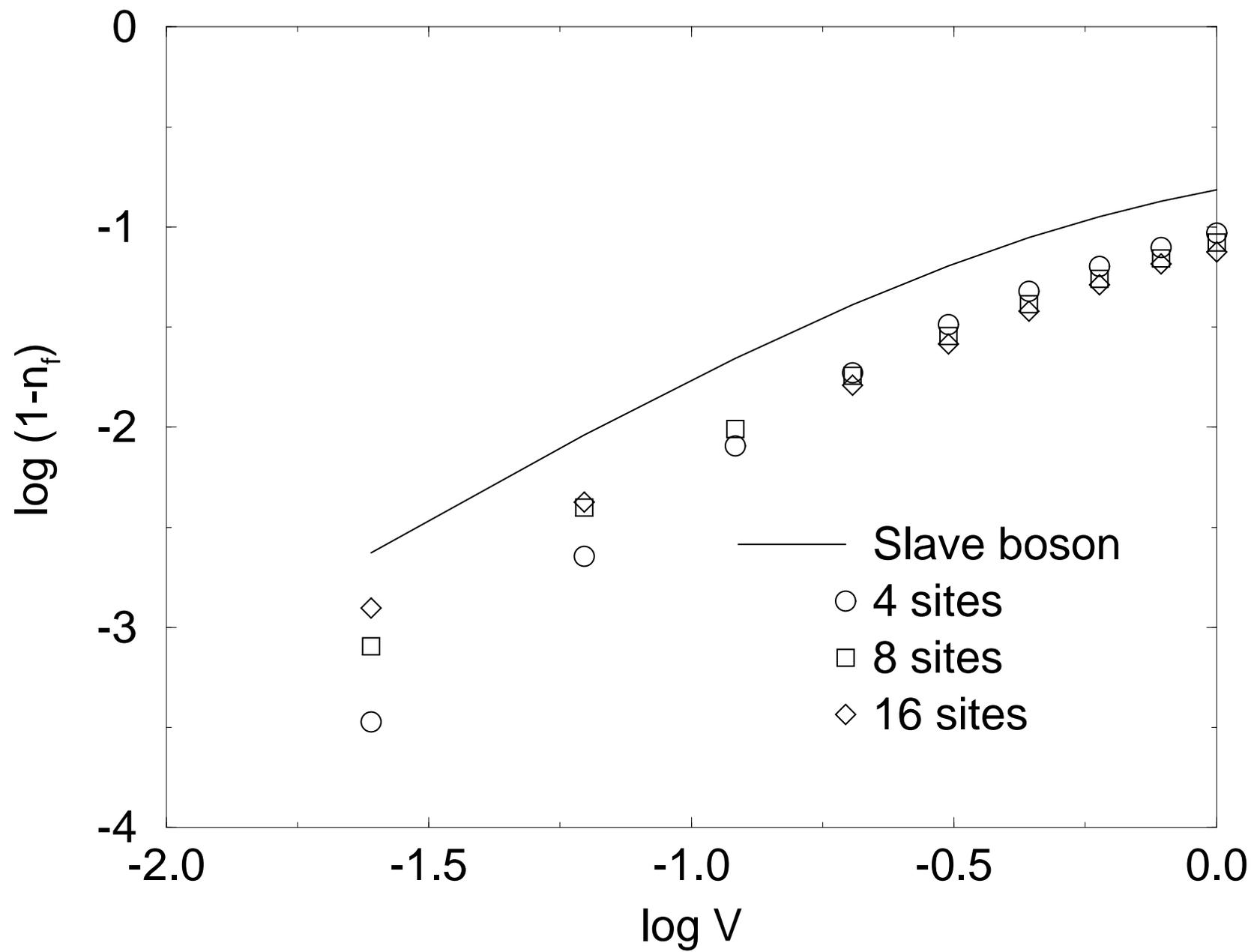
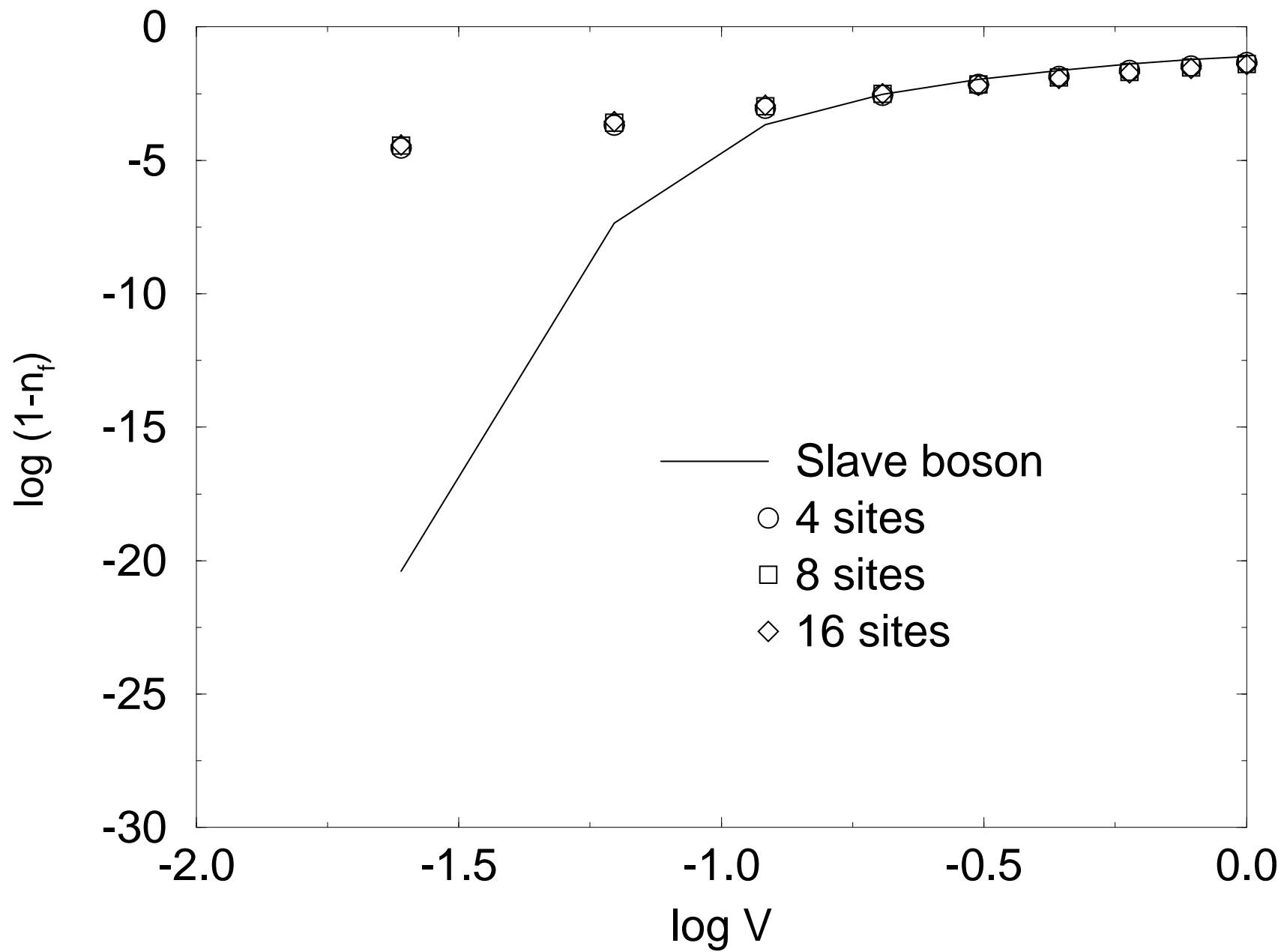
\end{document}